02 February 2019

# Polymer Nanocomposites having a High Filler Content: Synthesis, Structures, Properties, and Applications

Christian Harito[1,2,3], Dmitry V. Bavykin[1], Brian Yuliarto[2,3], Hermawan K. Dipojono[2,3], Frank C. Walsh[1]

[1] Energy Technology Research Group, Faculty of Engineering and Physical Sciences, University of Southampton, SO17 1BJ, Southampton, U.K.

[2] Advanced Functional Materials (AFM) Laboratory, Engineering Physics, Institut Teknologi Bandung, 40132, Bandung, Indonesia

[3] Research Center for Nanosciences and Nanotechnology (RCNN), Institut Teknologi Bandung, 40132, Bandung, Indonesia

**≈ 13 000 words, 4 tables and 11 figures**

**Contents**

**Abstract**

The recent development of nanoscale fillers, such as carbon nanotube, graphene, and nanocellulose, allows the functionality of polymer nanocomposites to be controlled and enhanced. However, conventional synthesis methods of polymer nanocomposites cannot maximise the reinforcement of these nanofillers at high filler content. Approaches to the synthesis of high content filler polymer nanocomposites are suggested to facilitate future applications. The fabrication methods address design of the polymer nanocomposite architecture, which encompass one, two, and three dimensional morphology. Factors that hamper the reinforcement of nanostructures, such as alignment, dispersion of filler as well as interfacial bonding between filler and polymer are outlined. Using suitable approaches, maximum potential reinforcement of nanoscale filler can be anticipated without limitations in orientation, dispersion, and the integrity of the filler particle-matrix interface. High filler content polymer composites containing emerging materials such as 2D transition metal carbides, nitrides, and carbonitrides (MXenes) are expected in the future.

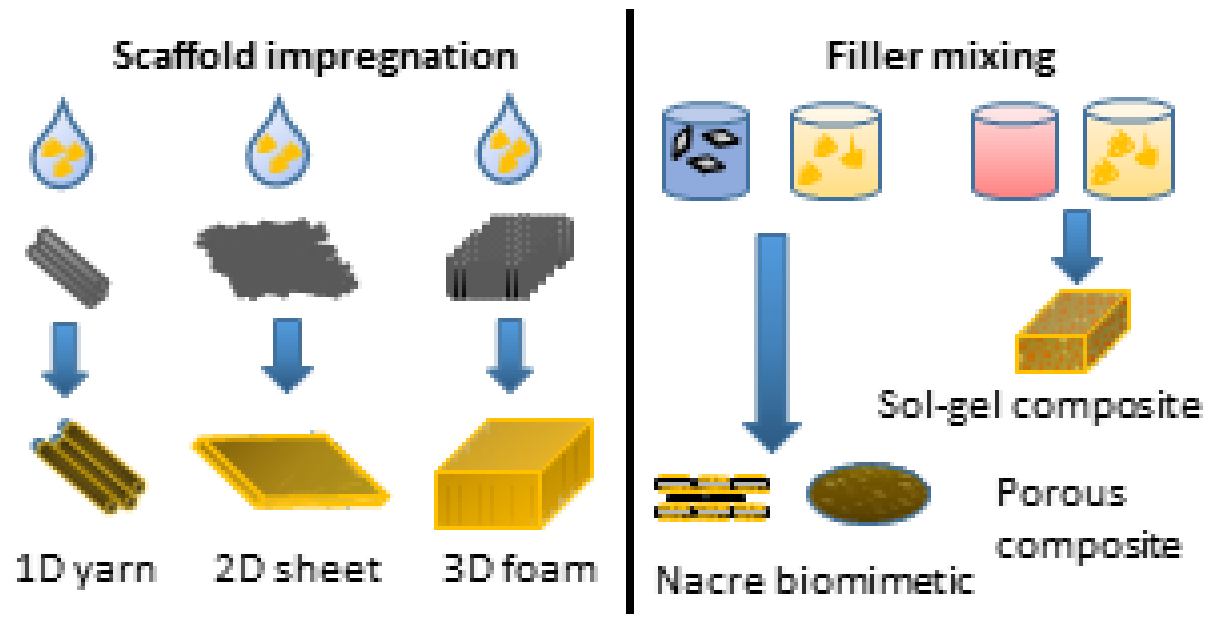

Graphical abstract: Approaches to the synthesis of high filler content polymer composites

## 1. Introduction

Polymer composites consisting of polymer and filler usually provide a better performance than pure polymers. The properties these composites are often dominated by the filler. Particles can be categorized as macro-, micro-, and nano-fillers depending on their size. The reinforcement of nano-fillers is significantly better than micro- and macro-fillers at the same filler loading [1]. This improvement is due to the large degree of contact between nano-filler and polymer, known as the "nano-effect" [1]. Putting a high content of nano-filler (e.g., carbon nanotube, graphene, or nano-$TiO_2$) may provide the polymer with new properties such as high electrical conductivity, refractive index, dielectric properties, mechanical properties, and unique response to certain stimuli such as pH, light, thermal, and magnetic fields. A 'high filler content' is considered in this review paper as 'containing at least 50 wt% filler'.

Carbon nanotube fillers have provided much promise due to their high strength, modulus and electrical conductivity. The Young's modulus and tensile strength of individual multi-walled carbon nanotube is 270-950 GPa and 11-63 GPa, respectively, measured by tensile testing in the nanotube length direction [2]. Due to their high strength and low density, CNTs present the best candidate for space elevator, which is a concept from NASA to reduce the cost of transporting material to space. This ambitious application has challenged researchers to grow the longest nanotube which has reached 550 mm in length [3]. However, the breakage of the first nanotube in nanotubes bundle leads to a great reduction on its mechanical properties [4]. Slip may occur between individual nanotubes surfaces when bundled without binder [5]. Polymer can be used as a binder to hold the fibres in unidirectional position and efficiently transfer load between fibres. High content of filler with adequate interfacial bonding with polymer matrix is needed to maximise the properties of filler which later can be used for advanced applications such as this space elevator. As an example, the strongest biological material (*i.e.*, limpet teeth, tensile strength of 3.0-6.5 GPa) is made from ≈80 vol% of aligned goethite nanofibres within a chitin matrix [6]. Many synthesis methods, such as solution processing [7–11], extrusion [12,13], and calendering [14] produce the polymer nanocomposites of random orientation and low nanotube content. Synthesis methods, that are able to produce a composite having a high filler content, need to be studied to meet the increasing demands on modern polymer nanocomposites.

Graphene is another exceptional nano-filler, which is one-atom thick carbon sheet with $sp^2$ hybridization, possessing an extraordinary electrical and mechanical properties [15]. Owing to its high electrical conductivity (up to $6 \times 10^3$ S $cm^{-1}$), incorporation of graphene within lightly cross-linked polysilicon (e.g., silly putty) will gives super sensitive electromechanical response that is able to detect pulse, blood pressure, and even small spider footsteps [16]. A single graphene sheet has a Young's modulus of 1 TPa and intrinsic strength of 130 GPa measured by nanoindentation in atomic force microscopy [17]. Utilisation of graphene for bulk nanocomposite often faces several problems such as poor dispersion [18], weak interfaces between graphene and binder [19] and brittleness will significantly increases [20]. Recently developed bioinspired nacre-mimetic synthesis has successfully produced dispersed graphene nanocomposite with content of more than 70 wt% while maintaining its elasticity [21]. Nacre has a "brick" and "mortar" structure consisting of 95 vol% inorganic $CaCO_3$ aragonite plate as a "brick" and 5 vol% organic "mortar" chitin and protein [22]. This material has a high toughness due to well distributed "brick" filler and strong interfaces between filler and polymer. By mimicking "brick" and "mortar" structure, nacre-inspired graphene-chitosan nanocomposites has tensile strength and toughness of 526.7 MPa and 17.7 MJ $m^{-3}$, which is 4 times and 10 times higher than natural nacre, respectively and electrically conductive [23]. This nature inspired design is suitable to create composite with high gas [24] and fire resistant [25]. A <0.1 mm thick heat shield can be made from nacre-inspired montmorillonite nanosheet-poly(diallyldimethyl-ammonium chloride) [25]. This ultra-thin high filler content nanocomposite is able to protect silk cocoon, which positioned 8 mm behind the composite, from *ca.* 2000 $^{o}C$ gas burner exposed to the other side of the composite.

An environmental friendly nano-filler (e.g., nanocellulose) has attracted increasing attention in recent years due to its low cost, renewable precursor, and biodegradability [26]. Nanocellulose can be extracted from natural resources, including algae [27], sea animal [28], and plant biomass [29,30]; it can also be synthetically grown from bacteria (e.g., *Acetobacter* species) [31,32]. The tensile modulus of single cellulose nanofibres was found to range from 100 to 160 GPa, which obtained by Raman spectroscopy [33] and X-ray diffraction [34–37]. However, a high nanocellulose content (>30 vol%) is needed to produce significant reinforcement of the polymer [26]. Human bone typically consists of ≈60-70% inorganic bone mineral and ≈30% collagen [38]. To ensure successful implantation and bone regeneration, replication of natural bone is needed [39]. The synthesis of high filler content polymer nanocomposites scaffold is essential to achieve better bone implant, high-performance environmental friendly composite, maximum mechanical and thermal reinforcement of nanofiller. Due to the many potential applications, this review will concentrate on synthesis of polymer nanocomposite with high filler content, their properties, existing and future applications.

## 2. Synthesis of high filler content polymer nanocomposites

Synthesis of polymer nanocomposite can be divided into several categories depending on a polymer or filler viewpoint. From a polymer perspective, there are three categories which are solution mixing, melt compounding, and *in-situ* polymerisation. In solution mixing, a disperse solution of nanofiller is mixed together with polymer solution. After achieving a homogeneous dispersion of nanofiller in polymer, evaporation of solvent is needed to leave nanofiller intact with the polymer. This method is simple and typically only requires relatively low temperature compare to melt compounding. Nonetheless, the method necessitates an appropriate solvent to adequately disperse the nanofiller and to dissolve the polymer well. Melt compounding is a blending process of nanofiller and polymer melt to manufacture polymer nanocomposite. This method usually requires high temperature to melt the polymer. Although melt compounding needs a high processing temperature, it does not require solvent as an intermediate between nanofiller and polymer. In the *in-situ* polymerisation, nanofillers are directly mixed with monomers solution and disperse while the polymerisation occurs thus reducing the fabrication time of polymer nanocomposites. However, nanofiller can hinder polymerisation.

Since this review focusing on high filler content, the synthesis method was classified from a filler perspective. The composite fabrication can be pursued by mixing dispersed filler and polymer solution (filler mixing) or by polymer impregnation of tailored filler scaffold (scaffold impregnation) [40]. Filler mixing is faster than scaffold impregnation which requires two steps: scaffold fabrication followed by polymer infiltration. However, the resulting product is usually thinner or less dense than scaffold impregnation due to the dispersion limit of filler. In scaffold impregnation, the geometry and morphology of filler can be tailored. It is important to ensure the process of polymer infiltration. The quality of polymer infiltration indicates the performance of composite thus the viscosity of polymer and its adsorbance to the filler should be controlled. Nanocomposites based on one (1D), two (2D), and three (3D) dimensional scaffolds and several mixing methods which able to produce high filler concentrations were represented in Figure 1. In this review, 1D fibre or yarn scaffold is defined as a composite which is significantly longer in only one direction or x axis than other directions such as y and z axis. 2D sheet or film has a significantly small thickness (z) compared to its length (x) and width (y). The 3D bulk composite has a shape in which the length (x), width (y), and thickness (z) cannot be negligible.

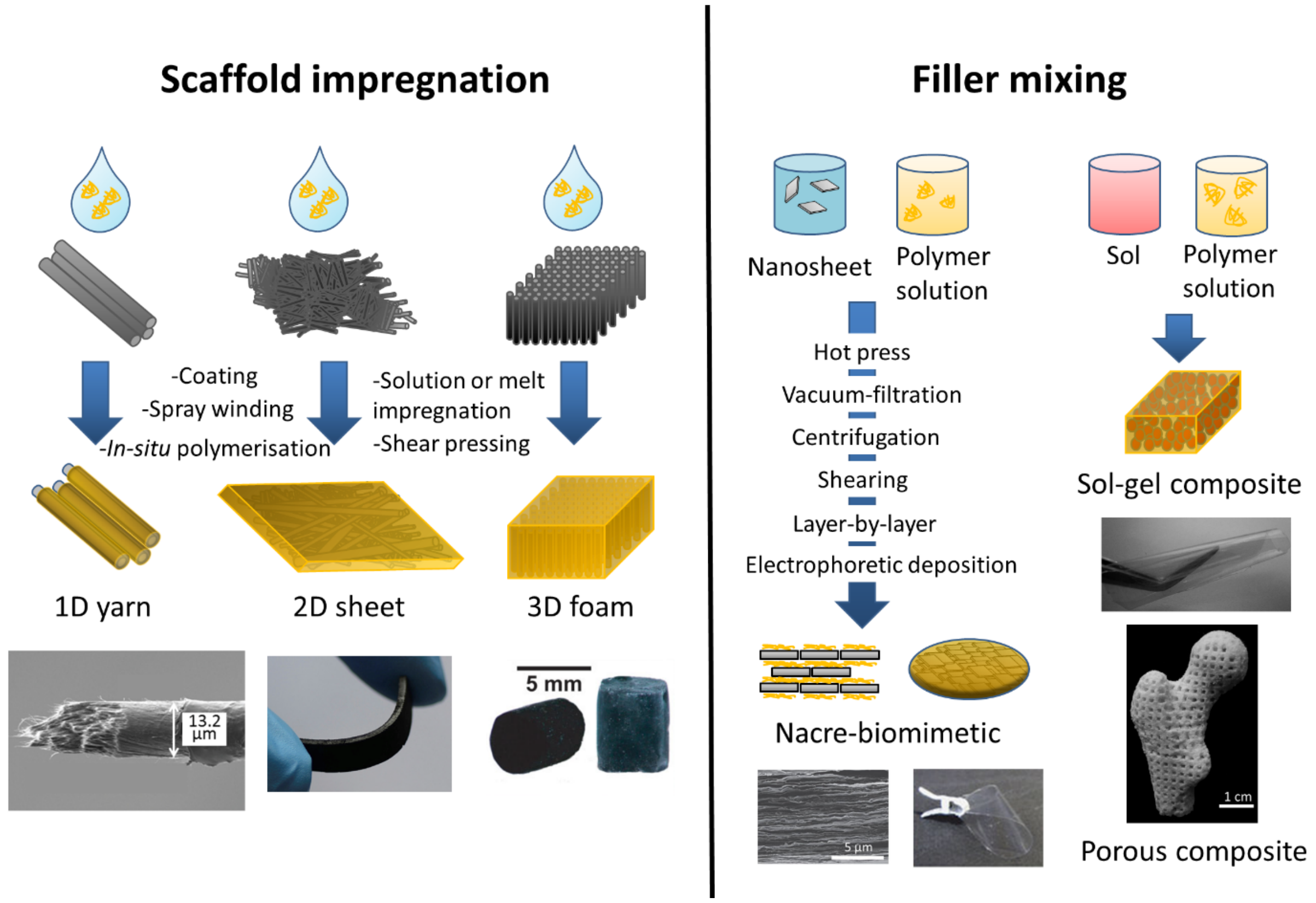

Figure 1. Approaches to synthesise high filler content polymer composites. Insets are taken from references [41–45]. Reprinted from Ref. [42] © 2016 with permission from Elsevier. Reprinted from Ref. [43] with permission from WILEY. Reprinted by permission from Springer Nature Costumer Service Centre GmbH: Springer Nature, Nature Communications, Ref. [44] © 2015. Reproduced from Ref. [45] with permission from The Royal Society of Chemistry.

## 2.1. Scaffold impregnation

### 2.1.1. 1D yarn

The race to create high performance fibres began in the 1960s, when DuPont synthesised aramid (aromatic polyamide) fibres, names Kevlar [46]. It was followed by ultra-high molecular weight polyethylene fibres (e.g., Spectra, Dyneema), liquid crystalline polyoxazole such as Zylon, and carbon fibres (e.g., polyacrylonitrile (PAN) and pitch based carbon fibres). Owing to their high mechanical properties, carbon nanotubes, CNTs have the potential to be next generation of high performance fibre. Instead of carbon nanotubes, graphene nanosheets can also be tailored into 1D in the shape of ribbon (rectangular cross section) or fibres (circular cross section). Methods to create one-dimensional (1D) carbon nanotube and graphene scaffolds can be categorised into solution spinning and solid-state spinning in Table 1.

Table 1 Synthesis and properties of CNT and graphene fibre scaffold. Adapted from [47] and [48]

| Solution spinning | | | | | | |
|---|---|---|---|---|---|---|
| Spinning technique | Materials | Post treatment | Mechanical properties | | | Electrical conductivity ($S\ m^{-1}$) |
| | | | Modulus (GPa) | Strength (MPa) | Elongation (%) | |
| Aqueous disperision coagulated in NaOH-methanol [49] | Graphene oxide | - | 5.4 | 102 | 6.8–10.1 | - |
| | Reduced graphene oxide | - | 7.7 | 140 | ≈ 5.8 | $2.5 \times 10^4$ |
| Aqueous dispersion coagulated in ethanol-water and $CaCl_2$ [50] | Graphene oxide | stretching | 11.2 | 364.4 | 6.8 | - |
| | Reduced graphene oxide | stretching | 11.2 | 501.5 | 6.7 | $4.1 \times 10^4$ |
| Aqueous dispersion coagulated in acetic acid-water and chitosan [51] | Graphene oxide | stretching | 22.6±1.9 | 442±18 | 3.6 ± 0.7 | - |
| Aqueous dispersion coagulated in ethyl acetate and $CaCl_2$ [52] | Graphene oxide | - | 6.2±1.3 | 139±8 | 11.1±1.5 | - |
| Aqueous dispersion coagulated in surfactant/$CaCl_2$ [53] | Large and small reduced graphene oxide | annealing | 135 ± 8 | 1080 ± 61 | ≈ 1.42 | $\approx 2.21 \times 10^4$ |
| Aqueous dispersion coagulated in acetone-ethyl acetate [54] | Large graphene oxide | stretching and annealing | 282 | 1450 | ≈ 0.6 | $0.8 \times 10^6$ |
| Graphene oxide hydrogel | Reduced graphene oxide | roll pressing | - | 404 | 2.25 | $5.7 \times 10^4$ |
| coagulated in ethanol [55] | Reduced graphene oxide | roll pressing and chemical reduction while wet | - | 102 | 85 | $3.55 \times 10^4$ |
| Chlorosulfonic acid dispersion coagulated in diethyl ether [56] | Reduced graphene oxide | annealing | 36.2 | 378 | 1.10±0.13 | $2.85 \times 10^4$ |
| Surfactant dispersion coagulated in water-PVA | SWCNT [57] | washing to desorb PVA | 15 | 150 | 3 | $10^3$ |
| | SWCNT [58] | washing and stretching | 40 | 230 | 0.9 | - |
| Surfactant dispersion coagulated in acid or base [59] | SWCNT | - | 12 | 65 | ≈ 1 | $1.5 \times 10^3$ |
| Sulfuric acid dispersion coagulated in water [60] | SWCNT | annealing | 120±10 | 116±10 | - | $5 \times 10^5$ |
| Chlorosulfonic acid dispersion coagulated in water [61] | SWCNT | stretching | 120±50 | ≈1000 | 1.4±0.5 | $\approx 2.9 \times 10^6$ |
| Commercial carbon fibre [48] | PAN based | - | 230-290 | 3530-7000 | 1.5–2.4 | $5.8\text{-}7.1 \times 10^4$ |
| | Pitch based | - | 343-588 | 2740-4700 | 0.7–1.4 | $0.9\text{-}1.4 \times 10^5$ |
| Solid-state spinning | | | | | | |
| Spinning technique | Materials | Post treatment s | Mechanical properties | | | Electrical conductivity (S/m) |
| | | | Modulus (GPa) | Strength (MPa) | Elongation (%) | |
| Gas-phase CVD | MWCNT [62] | - | - | 100-1000 | ≈ 100 | $8.33 \times 10^5$ |
| | long DWCNT [63] | acetone densificat ion | 50 | ≈1100 | ≈ 3.5 | - |
| | DWCNT [a 64] | water densifica-tion and roll pressing | 91 | 3760–5530 | ≈ 8-12 | $2 \times 10^6$ |
| | MWCNT [65] | twisting | 180 | 190 | ≈ 1.7-2.5 | - |
| | DWCNT [66] | twisting | 8.3 | 299 | ≈ 5.5 | - |
| Vertical-grown CNT array spinning | MWCNT [b] [67,68] | twisting | 5-30 | 150-460 | ≈ 13 | 3.03 x $10^4$ |
| | MWCNT [69] | methanol densification | 37 | 600 | 2.45 | - |
| | long MWCNT [c] [70] | twisting | 330 | 1910 | ≈ 7 | 4.17 x $10^4$ |
| | long DWCNT [d] [71] | twisting | 100-263 | 1350-3300 | ≈ 9 | 5.95 x $10^4$ |

[a] tested at a gauge length of 10 mm [b] 10 nm diameter and length 100 µm length

[c] 10 nm diameter and 650 µm length [d] 7 nm diameter and 1000 µm length

Carbon nanotubes and graphene fibre show promising result in new generation high performance fibre composites to replace carbon fibre. The highest modulus CNT and graphene fibres can achieve 330 GPa [70] and 282 GPa [54], respectively, comparable with PAN based carbon fibre, which has values of 230-290 GPa [48]. The CNT fibres can be stretched up to 7%, which is more than carbon fibre at 1.5-2.4%. These properties are still far short of the theoretical potential of carbon nanotubes. Voids within the fibres and slip between each nanotubes reduce mechanical properties. Post-treatment, such as twisting, may solve these problems via modified fibre alignment. As an alternative, nanotubes can be impregnated by polymer to strengthen bonding to the polymer matrix without sacrificing alignment. A good interfacial bonding between tubes and polymer is necessary to ensure adequate stress transfer from filler to matrix.

The ease of polymer infiltration can be determined by the permeability of CNTs, which represents the capacity of CNT fibre to transmit fluid. Permeability is determined by porosity, volume fraction of filler, tortuosity of the path, the viscosity of solution, applied pressure, the size of polymer molecules, surface area per unit volume, and infiltration distance [72]. Porosity of CNT fibres can be reduced by post treatment such as stretching to decrease the waviness of the CNT. Solvent needed to lower the viscosity of polymer solution thus it is able to infiltrate the CNT fibres. Applied pressure aids the polymer to infiltrate the fibres. However, higher atmospheric pressure exhibited lower degree of impregnation, due to compressed fibre bundle making the polymer with molecular size higher than the pore unable to penetrate. Longer duration of impregnation is needed to wet and infiltrate long CNT fibre with high surface area per unit volume.

Several types of polymer have been used to impregnate CNT fibres such as epoxy, polyvinyl alcohol (PVA), polyethylene imine (PEI), bismaleimide (BMI), and polyimide (PI). These impregnation methods produced composite with filler content >50 wt%. Comprehensive reviews about polymer/CNT fibre composite with low filler content has been covered elsewhere [73,74]. Water-soluble [75] and non-water soluble [76] low viscosity epoxy has deployed to impregnate twisted dry-spun CNT fibre of diameter ≈10 μm. CNT fibre was immersed in resin mixture of hardener and epoxy monomer for up to 5 hours followed by curing at high temperature ≈130 $^{o}$C. As expected, it was found that resin with lower viscosity was easier to infiltrate CNT and higher twisting rate of CNT was more difficult to be penetrated due to its compactness. Tensile strength and modulus of CNT/epoxy composites were higher than pristine CNT fibres while their elongations at break were slightly reduced. However, the epoxy was only able to penetrate ≈1 μm from CNT fibre surface. Several approaches are worth to try in the future such as vacuum impregnation, more impregnation duration, and CNT surface treatment to improve wetting of epoxy. Solvent can be used as alternative to aid polymer infiltration [77,78]. Densification of carbon nanotubes, directed by capillary force of solvent, may occur during solvent evaporation [79] which synergistically increase the mechanical properties of fibres. Ultrasonication may also assist polymer infiltration by loosening fibre bundles [80]. The vibration treatment was found beneficial to strengthen the composite fibre despite the fact that it may damage the outer surface of the fibre. Instead of twisting, polymer infiltration, and densification, the adhesion between polymer and filler should be taken into account to ensure high mechanical properties of composite fibres. Ryu, *et al.* [81] used catechol-based adhesive (polyethylene imine-catechol, PEI-C) mimicking the marine mussel *Mytilus edulis,* which is able to crosslink with CNT on heat treatment. The crosslinking of PEI-C at 120 $^{o}$C for 2 h significantly enhanced tensile strength from 0.91 ± 0.24 GPa to 2.2 ± 0.15 GPa and the modulus from 65 ± 17 GPa to 120 ± 23 GPa. The number of crosslinks was further improved by catechol oxidation with $Fe(NO_3)_3$ creating catechol–Fe(III) bonds, which improved the tensile strength up to 2.5 ± 0.31 GPa. Fe(II/III) ions are known to increase the mechanical properties of mussel protein by inducing the formation of catecholato-iron chelate complexes [82].

Other approaches of making CNT/polymer nanocomposites fibre such as polymer impregnation on CNT during [83] and before [84] fibre spinning have been studied. A schematic of CNT/polymer nanocomposite fibre synthesis is provided in Figure 2.

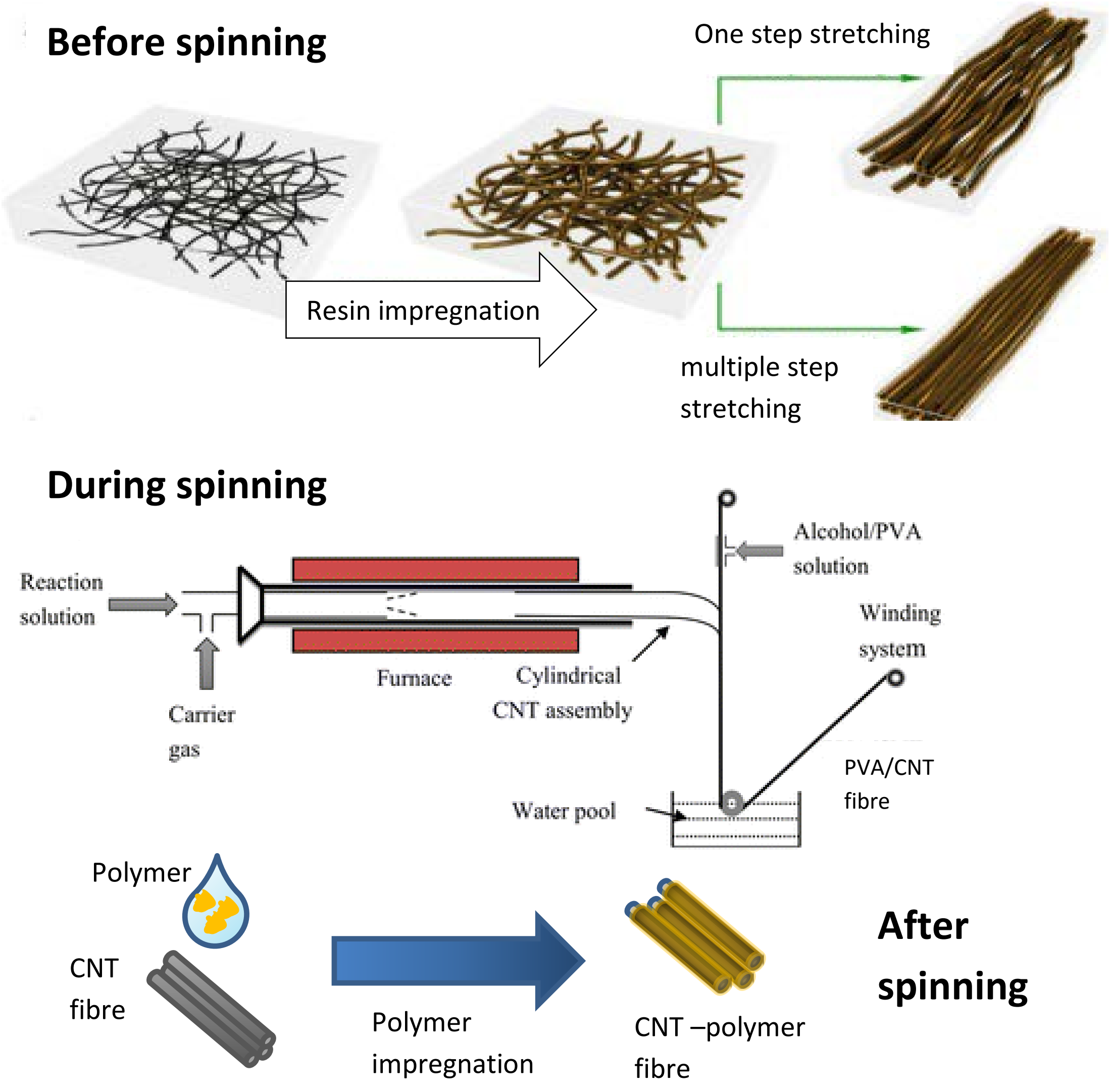

Figure 2. Schematic showing three different methods of polymer impregnation into CNT fibre (before [84], during [83], and after fibre spinning). Adapted with permission from Ref [83]. © 2016 American Chemical Society.

To deposit polymer on CNT during spinning, layer-by-layer deposition of polymer was performed on a vinylon wire substrate [83]. The wire was wetted with an alcoholic solution of polyvinyl alcohol (PVA). Densification occurred as CNT was deposited on wetted vinylon wire yielding one layer of PVA/CNT fibre. The supplying roller containing vinylon wire was then replaced by winding roller (CNT coated PVA/vinylon wire). After 5 layers deposition, the vinylon wire was dissolved in hot water at 90 $^{o}$C for 20 min and it was collapsed into ribbon-like PVA/CNT fibre with the width of 1.0-1.3 mm. The PVA/CNT fibre was wetted with PVA solution before use as a substrate. Spinning and rewetting of PVA/CNT fibre were repeated until the desired number of layers has been achieved. The weight fraction of CNT was controlled by adjusting concentration of PVA solution. The weight fraction of CNT was 67.5, 62.3, 57.7, 54.0, and 44.6 wt % for PVA concentration of 0.01, 0.05, 0.1, 0.2, and 0.4 wt %. PVA concentration was also crucial to control the thickness of the fibres which are 8.4, 5.0, 23.7 μm for 0.01, 0.05, and 0.2 wt% of PVA. If the deposited PVA solution is too low, densification of CNT is not optimum whereas PVA may not able to penetrate the CNT fibre at high concentrations. Alternatively, polymer can be used as a coagulant in wet spinning and used directly without washing [85]. The resulting CNT/PVA fibre with 60 wt% of CNT has around 100% elongation, modulus of 80 GPa, and tensile strength of 1.8 GPa.

CNT/polymer nanocomposite fibre can also be synthesised using impregnated 2D CNT preform [84]. The 2D CNT film was immersed in 1 wt% bismaleimide (BMI)/acetone solutions. CNT to BMI mass ratio was controlled by adjusting the volume of BMI solution. Adequate wetting of CNT (7:3 CNT/BMI mass ratios) must be achieved to ensure stretch-ability up to 30% without fracture. The impregnated films were stretched in two different ways. Firstly, those were stretched directly up to 30%. Secondly, the films were stretched multiple times until 30% with 5-10 min relaxation every 3% elongation. After stretching, the ribbon-like films were hot pressed to cure the BMI resin. The films with multiple stretching were more aligned than one-step stretching resulting in the tensile strength of 4.5–6.94 GPa, which is the highest tensile strength compared to the other techniques (e.g., CNT with the polymer impregnation during and after spinning). As a comparison with the one-step stretching after BMI infiltration, the 2D CNT film was stretched to 30% before BMI solution impregnation. The tensile strength was poor due to non-homogenous deposition of BMI, which are 585–801 MPa and 3.83-6.31 GPa for BMI impregnation with and without CNT film stretching. Fabrication of 1D CNT/polymer nanocomposites fibre using impregnated 2D preform is the most promising method to create homogeneous polymer distribution on CNT fibre. Polymer with large molecular size will be difficult to penetrate as-spun CNT fibre. While impregnation during spinning either requires complex setup or polymer that can be used as good coagulant. Deformability might be an issue for impregnated 2D scaffold. Further research is needed to explore the possibilities of tailoring impregnated 2D scaffold into long and thin fibre or braided into complex structures.

### 2.1.2. 2D sheet

2D scaffold can be obtained from simple drop casting or vacuum filtration of filler suspension creating carbon nanotubes buckypaper [86], graphene [87], clay [88], cellulose nanopaper [89], bacterial cellulose gel [90]. 2D building blocks (e.g., graphene and clay) are laid down parallel to the surface while 1D building block such as carbon nanotubes are usually randomly oriented. Alignment of carbon nanotubes buckypaper can be achieved by application of strong magnetic field (10-30 T) [91,92] or rotating cylinder (shear rate 640-1200 $s^{-1}$) [93] during filtration of CNT suspension. Carbon nanotubes can be paramagnetic or diamagnetic depending on the helicity of the nanotube, the field direction, the radius of the CNT, and the position of the Fermi energy [94]. When strong magnetic field applied (10-30 T), CNT shows anisotropic magnetic responses which align the CNT to the direction of magnetic field. The requirement of such strong magnet is not convenient. The high shearing method may be a visible option for large scale production.

Spinning (e.g., dry, wet, and electrospinning) is a versatile method to synthesise aligned or random 2D scaffold from 1D building block. Recently, electrospinning received increasing interest from researcher. Electrospinning is a versatile method which is able to produce polymer [95], carbon [96], and ceramic [97] nanofibres. High voltage is applied in electrospinning to create an electrically charged polymer solution or melt, which discharges from the tube tip and solidifies into fibres during its travel to the collector electrode. This polymer nanofibre can be converted into carbon nanofibres by carbonisation at high temperature (≈1000 $^{o}$C) in vacuum or in the presence of an inert gas shield to prevent oxidation. To spun metal oxide, low content of sacrificial polymer (10-30 wt%) with very high molecular weight is needed to be mixed with sol-gel precursor (e.g., zinc acetate, titanium butoxide). After spinning, the polymer is later decomposed into gas with heat treatment in air leaving only metal oxide in the form of nanofibre. The alignment of the fibres can be controlled by changing the collector electrode into rotating disc, parallel, ring [98], and concave curved electrode [99] which alter the electric field profile. The infiltration methods of 2D scaffold can be categorised into five methods in Figure 3.

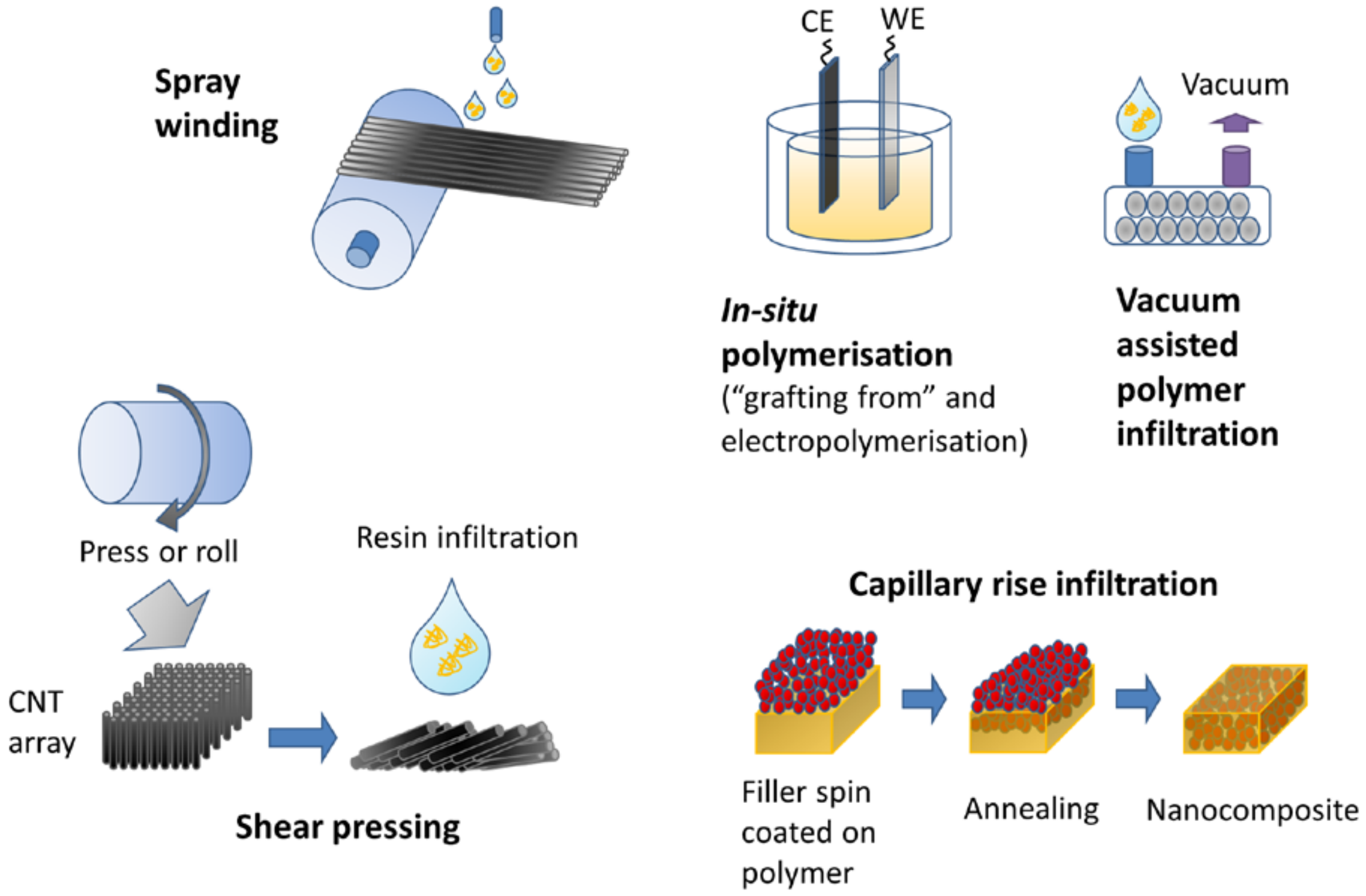

Figure 3. Infiltration method of 2D scaffold. Adapted from reference [100] with permission from The Royal Society of Chemistry.

The method to produce conventional fibre/polymer composites (e.g., vacuum assisted infiltration, spray winding, and *in-situ* polymerisation) can be applied to create nanofiller/polymer composites. Vacuum assisted infiltration is also known as vacuum assisted resin transfer moulding (RTM). Reinforcement (preform or scaffold) is placed into the mould and closed followed by polymer drawn into injection port with help of vacuum. Complex mould structure is often used to fabricate 3D nanocomposite structure from 2D nanofiller sheet. Nanofillers such as carbon nanotubes have such a small size enabling it to conform to the surface topography of the mould, thus complex structure can be synthesised [101]. The impregnation time of nanofiller can be estimated by adapting Darcy's law and Kozeny-Carman permeability equation with assuming behaviour of the polymer solution as Newtonian fluid [72,102]. Several factors affecting the permeability are similar to fibre which is porosity, volume fraction of filler, tortuosity of the path, the viscosity of solution, applied pressure, the size of polymer molecules, surface area, and infiltration distance.

Layer-by-layer deposition can be performed by spraying PVA solution on a rotating mandrel of CNT in which each rotation produce one layer of CNT/PVA composite [103]. The weight fraction of CNT was controlled by adjusting the concentration of PVA solution while keeping the same rotation speed. This spray winding method is able to reinforce pristine CNT fibres with polymer loading as low as 20 wt%. Instead of PVA solution, spray winding can also be used for BMI [104] and epoxy [105] resin by diluting in suitable solvent (e.g., DMF for BMI, acetone for epoxy). Post treatment such as stretching and heat treatment/hot press often used to increase the CNT alignment, reduce the porosity, and fully cure the resin.

Some monomers such as methyl methacrylate [106], styrene [107], nylon-6 [108], and conductive monomers [109] can be polymerised on the surface of modified nanofiller. First, the surface of nanofiller is functionalised with initiator such as 2,2'-azobisisobutyronitrile (AIBN) [106], Li/$NH_3$ [110], caprolactam [111], or oxidising agent for conducting polymer [112]. Then, monomers are added to the suspension to start the polymerisation and grow from the surface of nanofiller. This *in-situ* polymerisation method is called "grafting from" because the polymer grows from the surface of filler. Since the discovery of conducting polymer that lead to a Nobel prize in 2000 [113], controlled polymerisation using applied potential become an intriguing option. Conductive filler such as carbon nanotubes and graphene sheet can be used as an electrode. The electropolymerisation potential and efficiency are restrained through monomer concentration, the type of solvent, and pH of the solution [114].

Recently, novel methods have been introduced such as shear pressing and capillary rise infiltration. In shear pressing method, an array of CNT around 1 mm height was synthesised by chemical vapour deposition and oriented by pressing at an angle of 35 $^{o}$ to the substrate [115] or rolling [116], producing flattened and horizontally aligned CNT film preform. Pressing apparatus in Figure 4a was needed to fabricate consistent pressing angle and shearing force on the CNT array. The flattened CNT preform was carefully peeled from the substrate using tweezer. The rolling method, which is shown in Figure 4b, in a "domino push" method. Firstly, CNT array was covered by microporous membrane and pushed with a cylinder to horizontal direction with constant force, which is similar to hand lay-up procedure in conventional composite fabrication [117]. Secondly, a membrane covered CNT buckypaper was peeled off from silicon substrate. Thirdly, the aligned CNT buckypaper separated from microporous cover by ethanol. The buckypaper preform can be stacked into several layer [118] or directly use for resin impregnation (Figure 4c). Hot press is needed for layered preform to remove excess resin.

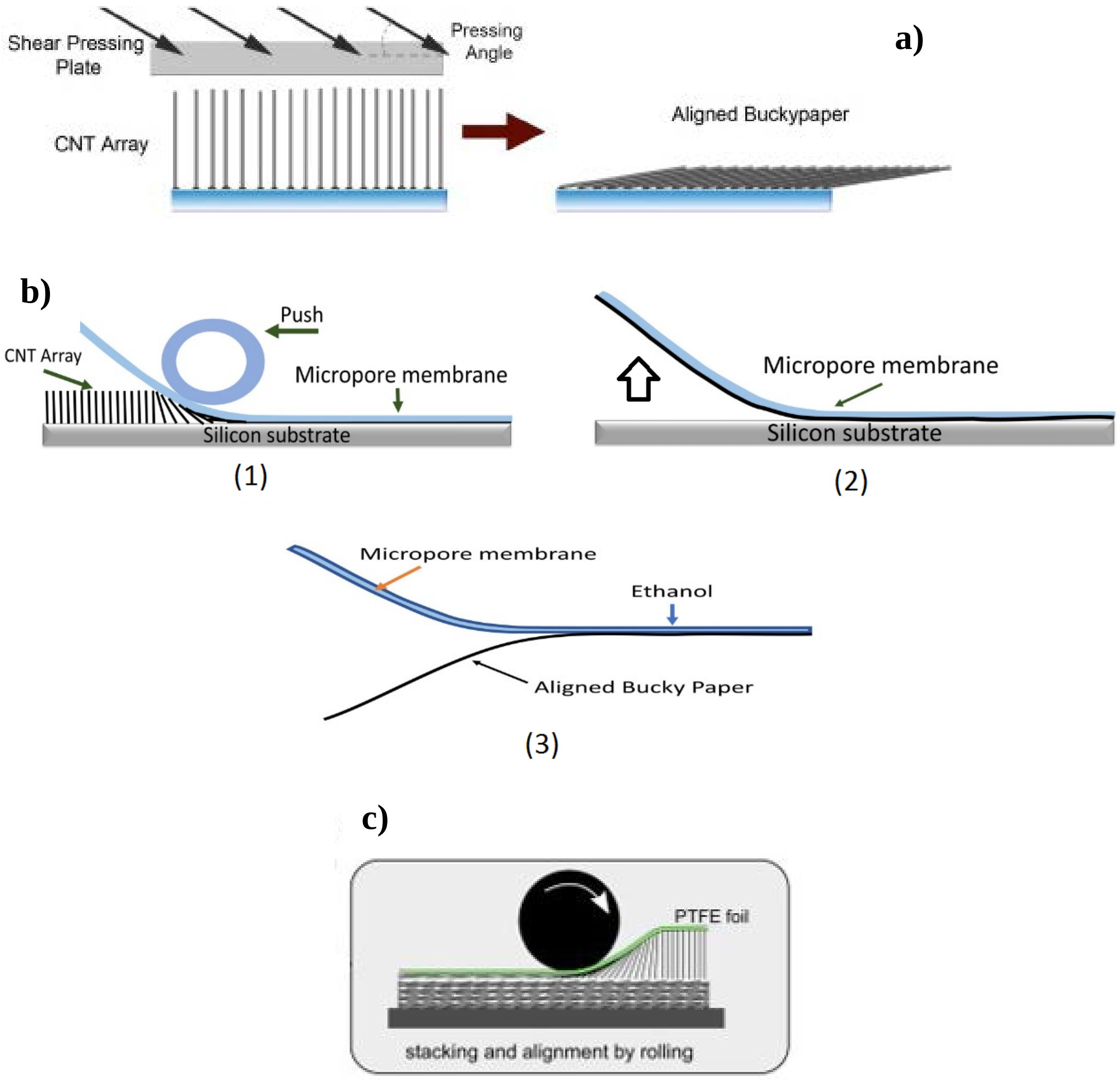

Figure 4. a) inclined angle shear press. Reprinted from Ref. [115] © 2010 with permission from Elsevier; b) domino push: (1) flattening the CNT array, (2) peeling membrane covered CNT from silicone substrate, (3) separating CNT buckypaper with membrane by ethanol. Adapted from reference [116]; c) stacking of aligned buckypaper Reprinted from Ref. [118] © 2015 with permission from Elsevier.

The capillary rise infiltration is based on Darcy's law of permeability [100]. Firstly, filler was spin coated on thermoplastic polymer such as polystyrene. Secondly, the polymer was heated above its glass transition temperature ($T_g$). The polymer became viscous when heated and infiltrate the filler by capillary force. The volume fraction of the filler can be controlled around 40 to 60 vol% depending on its packing density. This infiltration method is not limited to ceramic filler; it should be able to be applied to other 2D scaffolds. The advantages and disadvantages of each synthesis method are provided in Table 2.

Table 2 Advantages and disadvantages of composite synthesis methods using 2D scaffolds

| Synthesis method | Filler content | Advantages | Disadvantages |
|---|---|---|---|
| Spray winding | 50-80 wt% | -Large scale production<br>-Good alignment | Relatively complex apparatus |
| Shear press | 60-70 wt% | Good alignment | Small scale production |
| Capillary rise infiltration | 40-60 wt% | Simple apparatus | Limited to thermoplastic polymer |
| *In-situ* polymerisation | 5-70 wt% | -Polymerisation and composite fabrication occur at the same time<br>-A good interface between filler and polymer | Limited to certain type of polymer |
| Vacuum assisted polymer infiltration | 5-70 wt% | Capable of producing large and complex part | Thickness and filler fraction are difficult to control |

### 2.1.3. 3D array

A 3D ceramic scaffold can be synthesised by foaming, coating, templating, slip casting, granulation, tape casting, extrusion, and pulse current processing [119] as shown in Figure 5a.

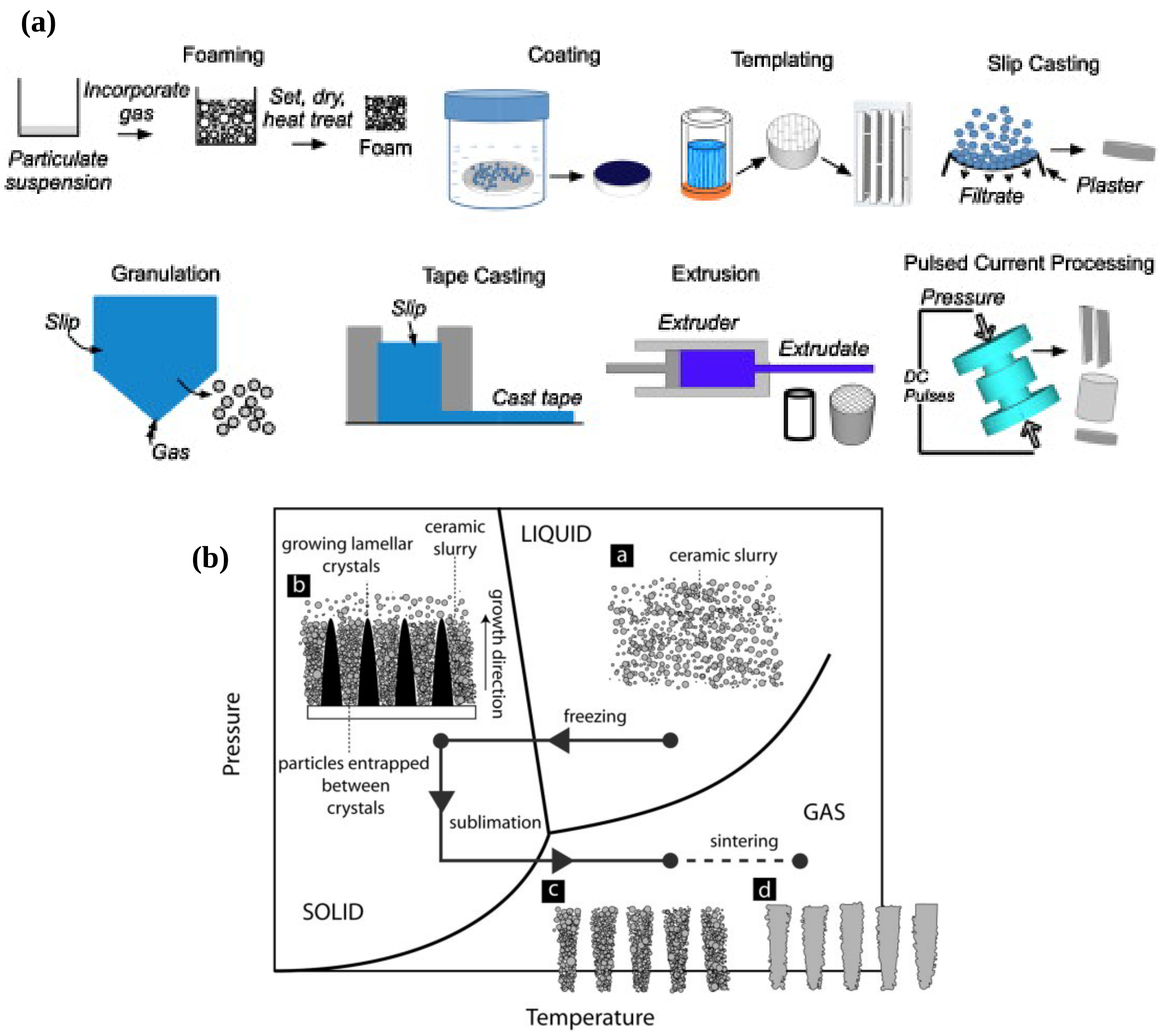

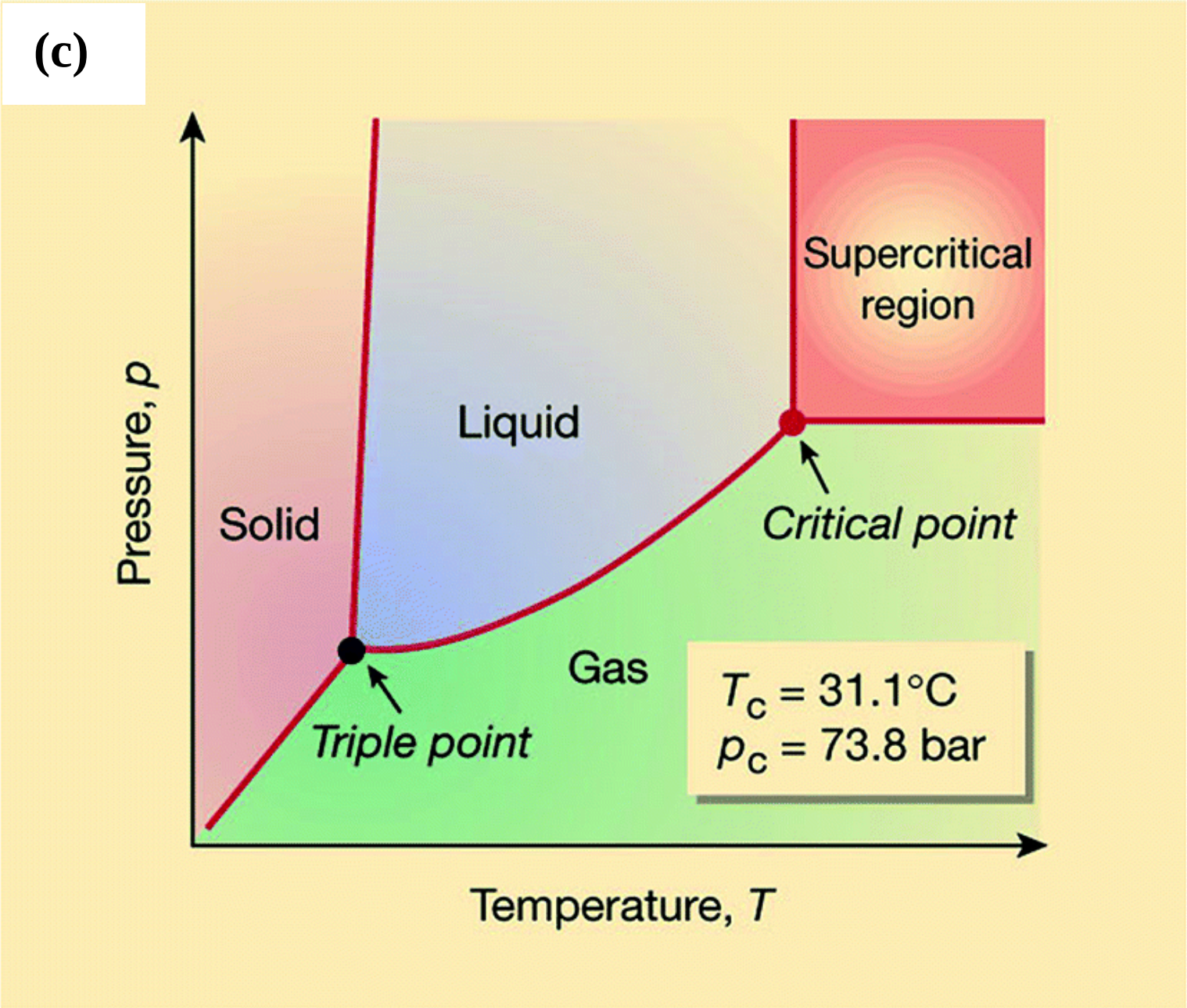

Figure 5. (a) Synthesis of a ceramic scaffold. Reprinted from Ref. [119] © 2014 with permission from Elsevier; (b) freeze casting/drying (ice templating) processes. Reprinted from Ref. [120] with permission from WILEY; (c) supercritical drying. Reprinted by permission from Springer Nature Costumer Service Centre GmbH: Springer Nature, Nature, Ref [121] © 2000.

Templating is one of the most versatile methods to control porosity of the ceramic scaffold in the range of 20-80 vol% which can be tuned by adjusting the amount of templates [119]. Soft template such as surfactant [122] and hard template (e.g., polymer beads, metal, metal oxide, carbon [123]) are commonly used as template. The bio-based molecules (e.g., protein) [124] and organism such as virus [125] affects the assembly and mineralisation process of the ceramic crystals, which can alter the porosity and surface area of the scaffold. Scaffold having lamellar structure can be synthesised by directional freeze casting/drying (ice templating) [120]. Freeze-drying consists of 3 main processes which are solidification/freezing, sublimation, and sintering depicted in Figure 5b. Firstly, liberation of particles from the ceramic slurry is entrapped between ice crystals during directional freezing. The higher cooling rate produces smaller pores size and higher solid content in slurry fabricates thicker ceramic walls [126].

Secondly, ice crystal is turned into gas phase creating porosity. Thirdly, sintering densifies and creates a solid lamellar structure. Contrary to freeze drying, which use low temperature process (Figure 5b), the supercritical drying involves high temperature and supercritical region (Figure5c), where the distinction between gas and liquid ceases to apply [127]. Among other fluid, $CO_2$ is the most common fluid for supercritical drying with critical temperature of 31.1 °C and pressure of 7.4 MPa, requiring lower energy to reach critical point than water at temperature of above 374 °C and pressure of over 22.1 MPa. Co-solvent can be used for sol-gel synthesis involving supercritical drying. A microporous solid, namely aerogel, can be formed by deploying supercritical drying of hydrogel in autoclave resulting solid 3D network without shrinkage or collapse. The high porosity of aerogel allows facile impregnation of polymer to coat its solid surface adding more functionalities. The emerging technology such as 3D printing [128,129] and electrospinning [130] can also be deployed to create 3D porous ceramic. With the porosity around 60-90 vol%, 3D printing and electrospinning are more suitable for polymer coated 3D network composites [128–130].

The strategy to create 3D ceramic scaffold can be applied to other materials as well. However, most of the 3D nanofiller scaffolds such as CNT sponge [131], graphene [132], and nanocellulose gel [133] are focused on improving light weight and porosity thus those are more suitable to make low filler content nanocomposites (0.1-1 wt%) or polymer coated 3D network composites. One must note recent development in hybrid nanocellulose/carbon nanostructures is able to enhance the dispersion limit of the carbon nanomaterials up to 75 wt% in water [134]. The high surface charge of nanocellulose (1400 μequiv $g^{-1}$) together with its strong affinity towards carbon nanomaterials enabled the stabilitation of carbon nanomaterials in water. In addition to a stabilising effect and providing more bonding options, the hierarchical architecture can be created by combination of 1D, 2D, and 3D filler enhancing its surface area [135–139]. With this hybrid method followed by solvent removal (e.g., freeze drying), the 3D scaffold of CNT or graphene with high filler content can be realised. High filler content nanocomposites can also be obtained by using a dense CNT array. An array of CNT is typically grown by chemical vapour deposition (CVD) at high temperature (600-1200 °C) using hydrocarbon gas as a carbon precursor, metal catalyst (e.g., Fe, Ni, Co) on a silicon or silica substrate [140]. The type of hydrocarbon gas and catalyst is crucial in determining the type of CNT (e.g., SWNT or MWNT) and the alignment of the array. Instead of aligned array, randomly oriented CNT sponge can be synthesised by CVD using ferrocene and 1,2-dichlorobenzene at 860 °C [141]. The synthesis of dense CNT array is performed by detaching the CNT array from the substrate followed by densification using solvent [142] or polymer solution [143]. Using n-hexane, the volume fraction of an aligned CNT array increase from 1 vol% to 20 vol% after the solvent dries, while randomly oriented CNT array increase to 4 vol%. The choice of solvent determines the resulting density of CNT array in this liquid induced collapse method. Futaba, *et al.* [144] were able to densify the CNT array up to ≈50 vol% of CNT using alcohol solution. The waviness of several millimetres long CNT array inhibits further densification of CNT. Surface tension, vapour pressure, and dipole moment of the solvent were identified as an important factor to determine densification level of CNT [145]. Dipole moment has more influence than surface tension and vapour pressure to create high level of densification [146]. Densification can also be performed without the use of solvent [147]. CNT was growth in low pressure environment and then exposed to ambient pressure. The pressure difference squeezed the CNT into denser array. However, it only shrunk to 26% of the original size which is less effective than the liquid induced collapse method (which resulted in 5% of the original size).

## 2.2. Filler mixing

By mixing dispersed filler suspension and polymer solution, nanocomposites can be synthesized and tailor into yarn by wet spinning, sheet, and bulky composite. For nanospherical filler, polymer composites can be created by sol-gel synthesis. Sol-gel is a method to synthesis metal oxides involving hydrolysis and condensation reactions [148] in alcoholic solution. The hydrolysis reaction of metal alkoxide precursor (e.g., $Si(OR)_4$, $Ti(OR)_4$) is controlled by adjusting water content and catalyst concentration (e.g., base or acid) and produces colloidal solution (sol). The reaction is followed by condensation reaction which creates interdigitated network of metal oxide (gel) and liberation of small molecule. This method is able to create particle with size of 1-2000 nm. Small particle size less than 20 nm is preferable for polymer nanocomposites synthesis since it is easier to disperse homogeneously. The particle can be synthesised with the presence of polymer (*in-situ*) or without (*ex-situ*). However, surface modified nanoparticle or *in-situ* synthesis is needed to create high filler content polymer composite up to 50 wt% without agglomeration [45,149,150].

Synthesis of polymer nanocomposites using 1D nanofiller (e.g., CNT, titanate nanotube) by filler or solution mixing usually produce a composite with low filler content (≤20 wt%). Agglomeration occurs at higher filler content worsen the properties of nanocomposites. For CNT, the viscosity of the polymer solution usually increases exponentially after 2 wt% addition of CNT, lowering its processablity [151]. This is not the case for other nanotubes such as titanate and halloysite clay nanotubes (HNT). The viscosity of chitosan solution was decreased with 30 wt% of HNT [152]. The viscosity started to increase after more than 50 wt% of HNT. Similar reduction was also observed in titanate nanotubes (TiNT) incorporation into polyamic acid solution [153]. Hydroxyl group of these ceramic nanotubes might selectively attract high molecular weight polymer lowering entanglement density of the polymer solution [154,155]. Although high filler content nanocomposites are processable using TiNT and HNT, homogeneous dispersion of nanofiller is still an issue that has to be resolved [156]. High filler content of these ceramic are suitable for the synthesis of porous nanocomposites tissue engineering scaffold due to its good biocompatibility [156,157] and low cost.

Bone tissue engineering scaffold must have sufficient mechanical properties while bone tissue regenerates in the pore and replaces the scaffold. Hydroxyapatite (HA), which consists of calcium phosphate, provides active sites for cell adhesion and proliferation thus it is commonly used as filler [158]. *In-situ* synthesis of fibrous hydroxyapatite showed homogeneous dispersion up to 65 wt% of HA in polyamide solution [159]. Chemical bond such as hydrogen bonding and/or carboxyl–calcium–carboxyl ($[-COO^-]-Ca^{2+}-[-COO^-]$) complex may formed between HA and polyamide preventing aggregation of HA. Alternatively, a 3D network composite scaffold can be achieved by freeze-drying of cross-linked hydrogel [160]. For example, graphene oxides (GO) and carbon nanotubes (CNT) disperse in polymer solution (e.g., polyvinyl alcohol, polyamic acid, polyacrylamide) or silane molecules may create cross-linked network [160]. The freeze-drying treatment of the cross-linked hydrogel followed by thermal annealing produces interconnecting network of porous GO or CNT/polymer composite up to 89 wt% of filler [161].

Nacre, which is inner shell layer of a mollusc, is an example of natural high filler content composites with a composition of 95:5 ratio of inorganic and organic. This composite has an intriguing “brick-mortar” structure with the inorganic part as “brick” and the organic part as “mortar”. By adapting this example from nature, 2D nanofiller such as graphene, layered double hydroxides, and montmorillonite clay can be synthesised by filler mixing to create nacre-like “brick-mortar” structure. A comprehensive review about nacre has been published around year 2012 [22,162]. It can be synthesised artificially by freeze casting, layer by layer, electrophoretic deposition, mechanical assembly. Freeze casting is an ice-templated unidirectional drying method to create 2D scaffold which has been discussed in section 2.1.2. Nanosheets such as clay and layered double hydroxide have surface charge in solution creating stable colloidal suspension. Some polymers bearing electrolyte group, are able to dissociate into positively charged polycation or negatively charged polyanion in aqueous solution. Multi-layer inorganic-organic composites can be synthesised by sequentially dipping a substrate (e.g., glass slide) in colloidal suspension and polymer solution with thorough rinsing in the interval, to make sure only one layer of nanosheets or polymer formed at a time. Each dipping produces ≈1 nm thick nanosheets layer and ≈10-30 nm of polymer layer, thus ≈100 bilayered film is needed to create 1 micron thick film which is time consuming. An electric field can be applied in electrophoretic deposition (EPD) to carry charged particles toward oppositely charged electrode. In layer-by-layer deposition (LbL), nanosheets and polymer must have opposite charge while in EPD the polymer can has same charge [163] or no charge [164]. LbL and EPD usually produces composite with ≈50 wt% of filler. Instead of polymer, monomer such as acrylamide can be used in EPD creating stacks of montmorillonite-acrylamide [165]. Polymerisation was conducted afterwards using UV light, yielding 95 and 5 wt% of montmorillonite and polyacrylamide, respectively. However, nacre-like composites synthesised by EPD has less ordered "brick-mortar" structure than layer-by-layer deposition. Synthesis of nacre-like composites by mechanical assembly is able to create thick film, large-scale, cost effective, and fast production. Nanosheets alignment is obtained using centrifugal, shear, gravitational, or pressure force in the mixture of nanosheets and polymer solution. Several methods of mechanical assembly are illustrated in Figure 6. Graphene oxide/polyacrylic acid composite up to 95 wt% of graphene can be produced using vacuum filtration [166]. Functional group of graphene oxide aids its dispersion in polyacrylic acid solution by creating hydrogen bonding. Graphene oxide was later chemically reduced in hydroiodic acid solution for 6 h to improve its conductivity.

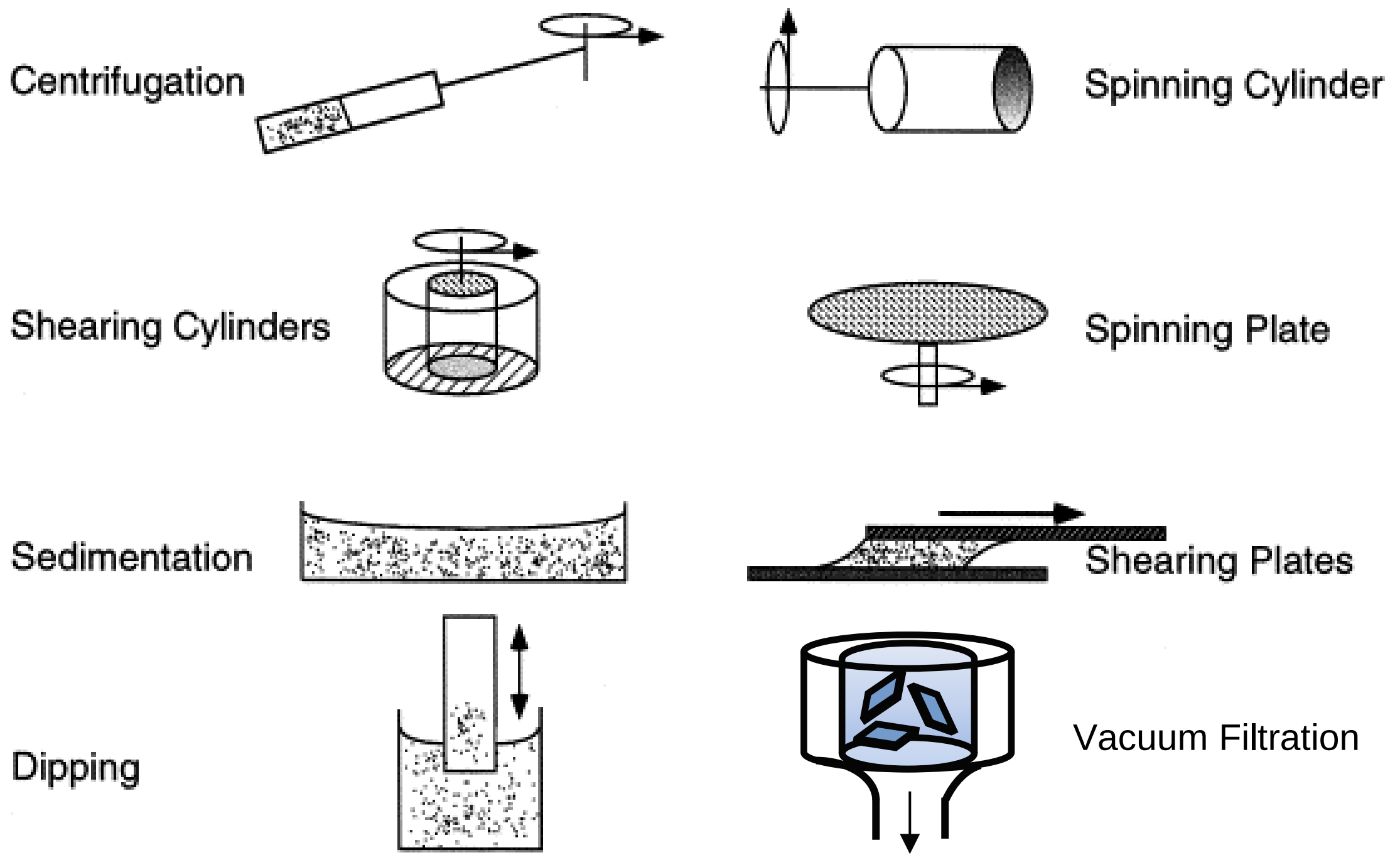

Figure 6. Mechanical assembly of a nacre-like composite. Adapted from Ref. [167] © 1999 with permission from Elsevier.

## 3. The properties of polymer nanocomposite with high filler content

This section highlights the features of polymer nanocomposite with high filler content such as mechanical, flammability, thermal and electrical properties compared to the low filler content composite. In ternary composite, the synergetic effect of hybrid fillers was seen up to high filler content (≥50 wt%) [168–171]. The synergism improved the mechanical [168–170], thermal [171] and electrical properties [172,173] of ternary composites compared to binary composites. However, most of high filler content ternary composite are still focused only on mechanical properties. Strategies to synthesise high filler content composites could be enable thorough studies of the thermal and electrical properties improvement due to the synergetic effect of hybrid fillers across a diverse range of filler contents.

### 3.1 Mechanical properties

To compare the mechanical reinforcement of filler in composite, several parameters must be fixed such as the properties of matrix and filler. The matrix stiffness plays a major role in the significances of filler reinforcements. Lijie *et al.* [174] showed that the reinforcement of carbon nanotubes (CNT) were decreased with the improvement of epoxy stiffness, which controlled by varying epoxy curing time. With the addition of 0.5 wt% CNT, the enhancement of the composite modulus was ≈200% down to 23%, or even negligible at the epoxy modulus of 0.15, 1.63, and 2.5 GPa, respectively. As observed with transmission electron microscopy (TEM), it was believed that the soft matrix has more physical contact at the interface of CNT and epoxy compared to hard matrix. In recent years, the use of graphene as mechanical reinforcement has exponentially increased due to its superior mechanical properties [17]. Here, the mechanical reinforcement of graphene for polyurethane, considered as soft matrix, are discussed to compare the effect of low and high content of filler.

The addition of a dispersed graphene oxide (GO) within thermoplastic polyurethane (TPU) increased the tensile modulus from 6.6 MPa to 9.8 and 28.7 MPa for 0, 0.9, 2.8 wt% of GO, respectively [175]. The reinforcement was even more prominent in functionalised GO/TPU, which reached up to 69.1 MPa tensile modulus improvement at 3 wt% isocyanate functionalised GO [175]. Although the improvements were remarkable at low filler content, the modulus of GO/TPU can reached around 166 ± 9.7 MPa at 90 wt% GO, with the toughness of 3.3 ± 0.3 $MJ \cdot m^{-3}$ [176]. The 90:10 wt% of GO/TPU composite was synthesised using vacuum filtration creating nacre bio-mimetic structure of "brick" and "mortar". The vacuum filtration assembled the stacks of nanosheets with the TPU as glue between the nanosheets. The unidirectional orientation gives a huge benefit in the composite tensile properties. However, the modulus is still far away from the modulus of graphene, which is 1 TPa [177]. While strengthening the interface between filler and polymer increases the modulus of the composite, it sacrifices the toughness due to the brittle nature of the filler [178]. The crack deflection should be facilitated to improve the ductility of the nacre bio-mimetic composite whether by using other materials (e.g., $MoS_2$, CNT, montmorillonite) or by the synergetic effect of covalent and non-covalent bonding (e.g., hydrogen bonding, polar, $\pi$-$\pi$ interactions), in Figure 9 [21]. The covalent and non-covalent bonding act as crack bridging during fracture facilitating the composite to be stretched before ultimate failure, which is known as pseudo-ductile failure [21]. Recent strategies to achieve strong and tough nacre inspired GO/polymer are provided in the following Table 3.

Table 3. Recent developments in the strength, modulus and toughness of nacre inspired GO/polymer.

| Material | Tensile strength (MPa) | Modulus (GPa) | Toughness (MJ $m^{-3}$) | Crack deflection strategy |
|---|---|---|---|---|
| rGO/10,12-pentacosadiyn-1-ol (PCDO) (96:4) wt% [179] | 348.5 ± 12.0 | Not available | 8.5 ± 1.3 | Covalent bonding |
| rGO/CNT/ 10,12-pentacosadiyn-1-ol (PCDO) (95:1.25:3.75) wt% [180] | 374.1 ± 22.8 | Not available | 9.2 ± 0.8 | Hybrid with carbon nanotube and covalent bonding |
| rGO/ 10,12-pentacosadiyn-1-ol (PCDO) (97:3) wt% [181] | 439.1 ± 15.9 | Not available | 7.6 ± 0.5 | Ionic (Zn) bonding and covalent crosslinking |
| rGO/10,12-pentacosadiyn-1-ol (PCDO) (96:4) wt% [179] | 688.5 ± 17.0 | Not available | 16.6 ± 1.2 | Covalent bonding and $\pi$-$\pi$ interactions |
| rGO/10,12-pentacosadiyn-1-ol (PCDO) (96:4) wt% [179] | 944.5 ± 46.6 | Not available | 20.6 ± 1.0 | Covalent bonding and $\pi$-$\pi$ interactions |
| rGO/gelatin (70:30) wt% [182] | 902.1 ± 90.4 | 27.84 ± 2.24 | 17.83 | Mechanical annealing, hydrogen and covalent crosslinking |
| GO/chitosan (CS) (95:5) wt% [183] | 347.0 ± 19.3 | 10.2 ± 3.1 | 10.6 ± 0.3 | Covalent and hydrogen bonding |
| rGO/chitosan (CS) (95:5) wt% [183] | 526.7 ± 17.3 | 6.5 ± 0.3 | 17.7 ± 2.4 | Covalent and hydrogen bonding |
| rGO/chitosan (CS) (94:6) wt% [184] | 868.6 ± 40.6 | Not available | 14.0± 1.2 | Covalent, hydrogen, and ionic (Cu) bonding |
| GO/sulfonated styrene-ethylene/butylene-styrene (SSEBS) (90:10) wt% [185] | 158 ± 6.0 | Not available | 15.3 ± 1.5 | Hydrogen bonding and $\pi$-$\pi$ interactions |
| rGO/polyvinyl alcohol (PVA) (58:42) wt% [186] | 129.3 ± 12.2 | 2.02 ± 0.12 | 7.30 ± 1.08 | Interlayer hydrogen and covalent bonding |
| rGO/MMT/polyvinyl alcohol (PVA) (81:9:10) wt% [187] | 356.0 ± 15.5 | Not available | 7.5 ± 1.1 | Hybrid with montmorillonite, hydrogen and covalent bonding |
| rGO/polyvinyl alcohol (PVA) (92.7:7.3) wt% [188] | 327 ± 19.3 | Not available | 13.0 ± 0.7 | Quadruple hydrogen bonding by polydopamine capped GO |
| rGO/2-ureido-4[1H]-pyrimidinone (UPy) (85.5:14.5) wt% [189] | 325.6 ± 17.8 | Not available | 11.1 ± 1.3 | Quadruple hydrogen bonding by polydopamine capped GO |
| rGO/poly(acrylic acid-co-(3-acrylamidophenyl) boronic acid) (PAPB) (96:4) wt% [190] | 382 ± 12 | 4.31 ± 0.08 | 7.50 ± 0.40 | Hydrogen bonding and $\pi$-$\pi$ interactions |
| rGO/$MoS_2$/thermoplastic polyurethane (TPU) (86:4:10) wt% [191] | 235.3 ± 19.4 | Not available | 6.9 ± 0.5 | Hybrid with $MoS_2$ and hydrogen bonding |
| rGO/carboxymethyl cellulose (CMC) (85:15) wt% [192] | 475.2 ± 13.0 | Not available | 6.6 ± 0.3 | Hydrogen bonding and ionic (Mn) crosslinking |
| rGO/hydroxypropyl cellulose (HPC) (97:3) wt% [193] | 274.3 ± 8.7 | Not available | 6.7 ± 0.6 | Hydrogen bonding and ionic (Cu) crosslinking |
| GO/ polydimethylsiloxane-poly(glycidyl methacrylate) (PDMS-PGMA) (85:15) wt% [194] | 309 | Not available | 6.55 | Hydrogen bonding and physical crosslinking |
| GO/polydopamine (PDA) (90:10) wt% [195] | 170 | Not available | 5.6 | Hydrogen bonding, covalent crosslinking, and water content optimisation |
| rGO/polydopamine (PDA) (95:5) wt% [196] | 204.9 ± 17.0 | 6.1 ± 1.4 | 4.0 ± 0.9 | Hydrogen bonding and covalent crosslinking |
| GO/Poly (n-butyl acrylate) (PBA) (96.5:3.5) wt% [197] | 187 | 7 | 4.3 | Covalent grafting and hydrogen crosslinking |
| rGO/silk fibroin (97.5:2.5) wt% [198] | ≈300 | 26 | 2.8 | Hydrogen bonding and ionic (Al) crosslinking |

GO = graphene oxide; rGO = reduced graphene oxide

From Table 3, it can be concluded that the choice of polymer matrix have a strong influence on the strength and toughness of the composite. The high strength and toughness was shown in the rGO/gelatin [182] and rGO/chitosan [183] with both toughness around 17 MJ $m^{-3}$ without additional elements. Both polymer have enormous amine and hydroxyl groups providing sufficient crosslinking density and hydrogen bonding. This is consistent with recent computational analysis of the mechanical properties of graphene nacre composite [199,200]. This study stressed out the importance of optimum crosslinking density to improve the strength and toughness. Several strategies that can efficiently alter the crosslinking density includes graphene oxide functionalisation with polydopamine [188], $\pi$-$\pi$ interactions [185], and the utilisation of multivalent cationic ions (e.g., Zn, Mn, Al, Cd, Cu, Mg, Ca) [201]. Polydopamine functionalisation of graphene oxide increased the concentration of surface functional groups up to 4 times [189] enabling more crosslinking with other polymer such as polyvinyl alcohol (PVA) [188] and 2-ureido-4[1H]-pyrimidinone (UPy) [189]. The graphene oxide can be non-covalently functionalised with 1-pyrenebutyric acid N-hydroxysuccinimide ester (PSE) and 1-aminopyrene (AP) providing bridging between adjacent rGO nanosheets [202]. The crack bridging through $\pi$-$\pi$ interactions and covalent crosslinking with 10,12-pentacosadiyn-1-ol (PCDO) polymer resulted the highest tensile strength and toughness at 944.5 ± 46.6 MPa and 20.6 ± 1.0 $MJ.m^{-3}$, respectively [179]. The $\pi$-$\pi$ interactions can be also created by the incorporation of carbon nanotubes which acted as hybrid filler [180] and by using a styrene based polymer such as sulfonated styrene-ethylene/butylene-styrene (SSEBS) [185]. Furthermore, the improvement in tensile strength can be pursued by ionic bonding through multivalent cationic ions, which serve as crosslinking agent via metal–ligand coordinate bonds. Such ionic crosslinking is found in nature, in the jaws of marine polychaete *Nereis* and *Glycera*, improving its mechanical properties [203].

### 3.2 Flammability

Based on the thermal application, the incorporation of nanofiller can be divided into two purposes, which are decreasing the flammability and to increase the thermal conductivity. Heat release rate is one of the most important parameters to assess the flammability of a product and its fire hazard [204]. The heat release rate measures the rate of heat energy that is released by fire, which the higher value means faster heat transfer through a material. Clay has been profusely studied as flame retardant of polymer especially polyurethane foam, which is a combustible cushioning material [205]. The organically modified clay with quaternary ammonium salt was shown to decrease the peak of heat release rate (PHRR) by 16.82% and 31.13% with the addition of 5 and 10 wt% organoclay, respectively [206]. The type of the organic that used to modify the clay was also plays a role in flammability behaviour. The PHRR showed 15.9, 30.2 and 26.4% reduction for octadecyl trimethyl ammonium (ODTA), octadecyl primary ammonium (ODPA), and decanediamine (DDA) at 2 wt% organoclay, respectively [207]. However, the organic may react with fire, slightly reducing the time to ignition [208]. Instead of clay, the flammability reduction was also observed in layered double hydroxide nanocomposites [209,210]. The nacre inspired high filler polymer nanocomposites were commonly deployed as coating to reduce the flammability of polyurethane foam. The PHRR of polyurethane foam was decreased by 84.1% with the ≈1-2 micron thick coating of montmorillonite/carboxymethyl chitosan (MMT/CCS) at 50 wt% MMT via one layer of dip coating [211]. The approach to make even thinner coating (less than 100 nm) were made by layer-by-layer (LbL) coating [212,213]. The MMT was assembled with polyacrylic acid (PAA) and polyethyleneimine (PEI) to fabricate nacre composite reducing the PHRR of polyurethane foam by 42% via 10 bilayer of LbL coating [212]. By combining cationic layered double hydroxide (e.g, boehmite) and anionic clay (e.g., vermiculite) with PAA and PEI as glue, the PHRR was decreased by 55% with only one bilayer of LbL coating [213]. The alignment of stacked nanosheets created tortuous pathway for oxygen reducing the flammability [214].

### 3.3 Thermal and electrical conductivity

Contrary to the flame retardant, the application such as heat sink or heat transfer agent pursues the improvement of thermal conductivity. In non-metal materials, the thermal conductivity is generally dominated by phonons rather than electrons as heat carriers [215,216]. To ensure a good phonon conduction in composite, the interfacial thermal resistance, and the concentration and the geometric shapes of fillers should be considered in composite design [217]. The interfacial thermal resistance governs by the thermal contact resistance (TCR), which defines by the poor interfacial bonding between filler and matrix, and the thermal boundary resistance (TBR), which occurs due to differences in the physical properties of filler and matrix [218]. Such poor interface hampers the phonon scattering and vibration restricting the heat flow. The functionalisation may increase [219] or decrease [220] the thermal contact resistance depending on the functional group [221] and its influence to the filler physical properties [222]. The filler, as an additive to increase the thermal conductivity, is also crucial in controlling thermal conductivity of composite. Most of the researches on low filler content composites are focused on the percolation network of filler, in which the fillers within matrix are connected and creates pathway for phonon or electron to easily transfer the heat [217,223]. When the percolation network exists within the matrix, the thermal conductivity is exponentially increased, which usually occurs around 10-20 wt% of conductive filler [223]. A good dispersion, morphology and combination of fillers (e.g., one and two dimensional) become an important factors to achieve percolation. Meanwhile, the percolation at high filler content is assured even the composite thermal conductivity may close to the filler conductivity. The thermal conductivity increased from $10^{-8}$, 0.02, to 287 W. $m^{-1}$ $K^{-1}$ at <10 wt% (no percolation), 10-20 wt% (percolation occurs), and CNT/polyvinyl alcohol (CNT/PVA) fibre composite with 80 wt% CNT, respectively, in which full CNT fibre was measured at 456 W. $m^{-1}$ $K^{-1}$ [224]. Anisotropy in thermal conductivity occurred especially in few walled carbon nanotubes [225] hence re-orienting the fibre (e.g., stretching) in one direction improved the thermal conductivity [226]. Same anisotropy also happened in nacre inspired graphene/PVA composite where the ratio of lateral and perpendicular thermal conductivity ($k_{\parallel}/k_{\perp}$) is 380 [227]. The percolation phenomenon can be found as well in electrical conductivity of composite. Since the electron also act as carrier in electrical conduction, the weight fraction needed to achieved percolation is significantly lower, which usually starts from below 1 wt% of nanofiller [228]. The electrical conductivity increased from $10^{-9}$, $10^{-6}$, to 920 S. $cm^{-1}$ at 1 wt% (no percolation), 2 wt% (percolation occurs), and CNT/PVA fibre composite with 95 wt% CNT, respectively [229,230]. To sum up, the high filler content polymer nanocomposite provides superior mechanical [176], thermal [212,213,224], and electrical [229,230] properties with several magnitudes higher than low filler content polymer composite. However, the brittleness should be reduced by crack deflection strategies (e.g., using lubricant or synergetic bonding) and the anisotropy of filler can be optimised by the alignment of fillers such as stretching.

## 4. Factors affecting properties of polymer nanocomposites

To make high quality polymer nanocomposites having a high filler content, key problem in polymer nanocomposites must be identified. The key issues and common problem in developing polymer composites are dispersion, interface between filler and polymer, alignment of nanofiller, and quality of nanostructures [231]. Defects on nanostructures will also affect the properties of polymer nanocomposite. However, the quality of nanostructures is outside the scope of this review.

### 4.1. Alignment of filler

Some nanostructures such as titanate nanosheets [232], graphene [233], carbon nanotubes [234], boron nitride nanotubes [235] have anisotropic properties thus alignment of nanostructures is preferable to harness maximum benefit from nanostructural filler. For carbon nanotube, such alignment in polymer nanocomposite can be achieved by aligning the filler before polymer impregnation (e.g., fibre spinning and shear press) which have been discussed in previous chapter. It also can be forced by electric field, magnetic field, and mechanical stretching. However, the force alignment is not very promising since the alignment often only occurs on the outer layer [40].

Aligned nanosheets within polymer can be obtained by filler mixing (e.g., nacre mimetic synthesis). Nanosheets such as clay, graphene, titanate nanosheets tend to stack on top of each other forming aligned brick formation. Even with simple drop casting, the degree of orientation can be as high as 95% within composite with 23 wt% of polymer [44]. For graphene fibre or yarn, the post-synthesis alignment such as rolling can be deployed to flattened the fibre into ribbon like graphene fibre [55]. Due to ordered structure of graphene within ribbon like graphene fibre, the tensile strength and electrical conductivity were improved almost twice from 238 MPa and $3.08 \times 10^4$ S $m^{-1}$ to 404 MPa and $6.3 \times 10^4$ S $m^{-1}$, respectively. The influence of polymer on nanosheet orientation becomes prominent at low filler concentration (<50 wt% of filler). For montmorillonite (MMT), 8 wt% of MMT only has 10% of the degree of orientation within polymer [236]. The orientation can be altered by applying shear stress with high shear rate (>10 $s^{-1}$) on the composite. However, percolation of nanosheets occur at >8 wt% of MMT, making it insensitive to shear and exhibits no change in orientation angle.

The degree of orientation or anisotropy factor can be determined by small-angle X-ray scattering (SAXS). In-plane and symmetrical scan of SAXS provide orientation information in 2 directions which perpendicular and parallel to the surface, respectively. The peak intensity of scattered radiation is attributed to number of stacking nanosheets, oriented parallel to the surface (Figure 7a & 7b). The homogeneity of oriented nanosheets manifest in several points scanning across the composite samples (Figure 7b).

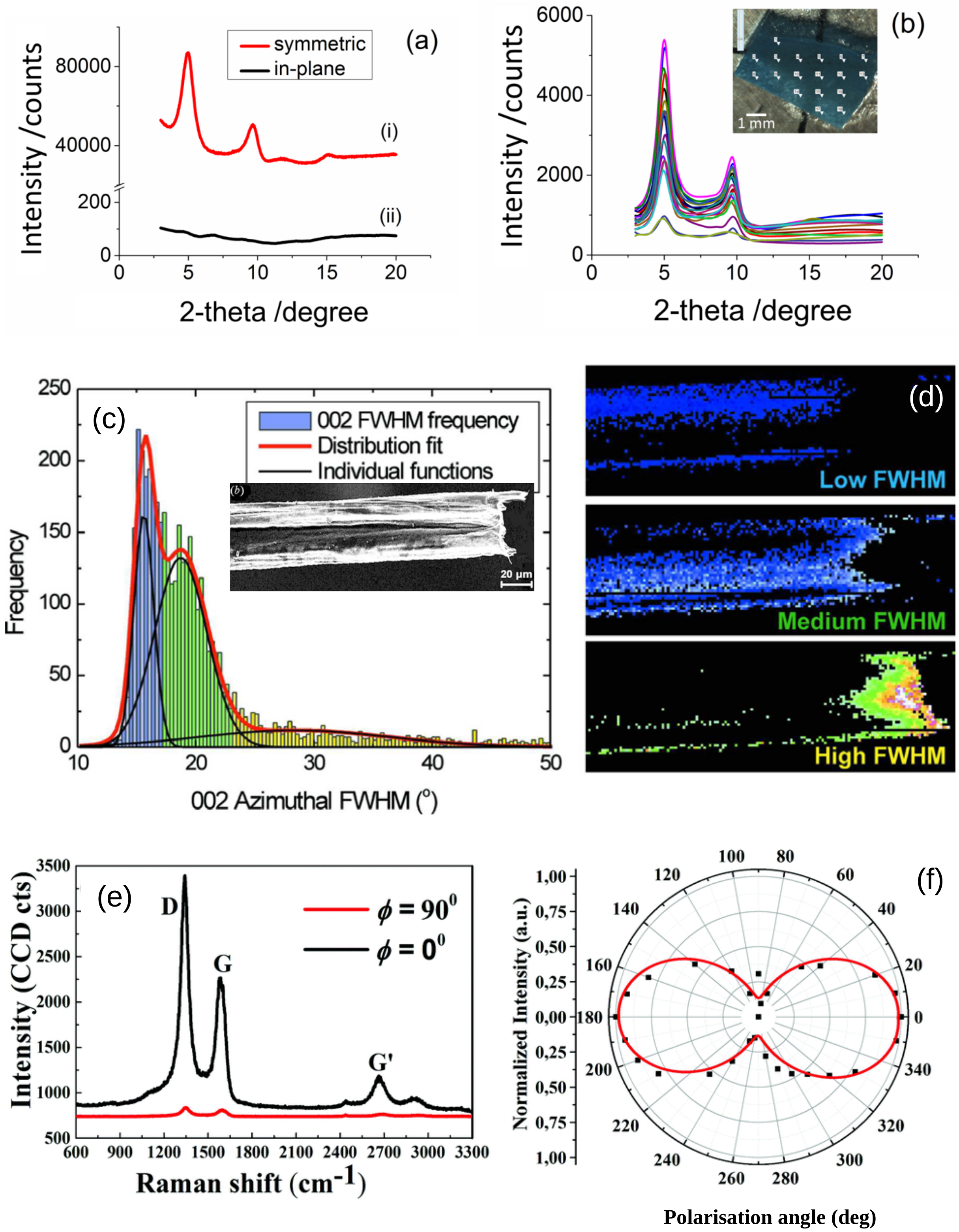

Figure 7. X-ray diffraction and polarised Raman spectroscopy to characterise the degree of orientation of filler in composite. (a) SAXS profile of titanate nanosheets (TiNS) by symmetrical (i) and in-plane (ii) scan; (b) SAXS mapping across TiNS/polymer composite. Reprinted from Ref. [237] © 1999 with permission from Elsevier; (c) histogram of the azimuthal

full width of half maximum (FWHM) of the 002 reflection (inset: SEM images of CNT fibre); (d) distribution of FWHM of 002 reflection [238]. Reproduced with permission of the International Union of Crystallography (https://journals.iucr.org/); (e) polarised Raman profile of aligned CNT; (f) G-band Raman intensity as a function of polarisation angle. Reproduced from Ref. [239] with permission from The Royal Society of Chemistry.

Intensity distribution profiles in the azimuthal angle ($\phi$), for particular $2\theta$ reflection along the Debye-Scherrer ring, also indicate the degree of orientation of the nanosheets [240]. The Debye-Scherrer ring corresponds to the cone projection in $\theta$ which allows for $\phi$ dependence The illustration and example of azimuthal scan in X-ray diffraction is provided in Figures 8a & 8b.

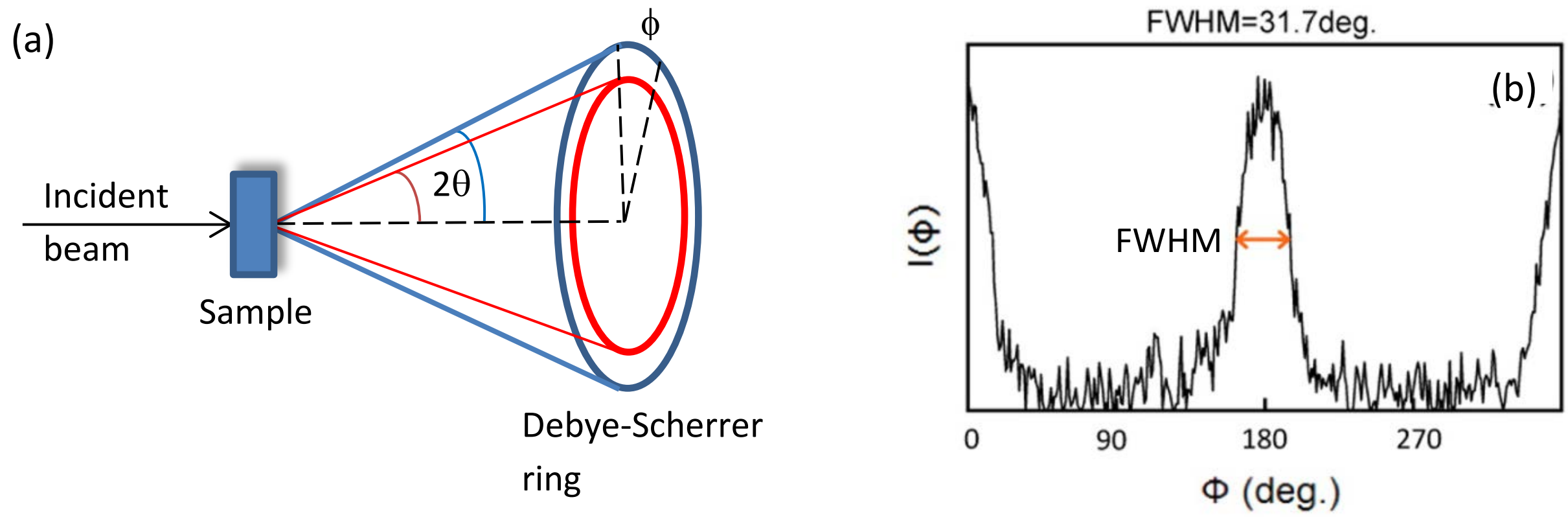

Figure 8. (a) Illustration of an azimuthal scan of X-ray diffraction along Debye-Scherrer ring ($\phi$ = azimuthal angle, $\theta$ = Bragg's angle); (b) an azimuthal intensity distribution along Debye-Scherrer ring of the reflection (060) of montmorillonite [240] (FWHM = full width at half maximum). Reproduced with permission from Ref [240] © 2013 American Chemical Society.

Figure 8b shows an example of the azimuthal scan of montmorillonite based nacre-mimetic composite film at (060) taken from the cross-sectional plane of the film. The azimuthal peaks at around 180° indicated the in-plane orientation of montmorillonite in the film. The degree of nanosheets orientation can be estimated semi-quantitatively using full width at half maximum (FWHM) of azimuthal peak or with the azimuthal angle and intensity by Herman's orientation factor (*f*). The calculation of degree of orientation using FWHM (Π) can be stated as follows

$$\% \, Degree \; of \; orientation \; (\Pi) = \frac{180-FWHM}{180} x100 \quad (3.1)$$

where FWHM is the full width at half maximum (FWHM) of azimuthal intensity. Herman's orientation factor (*f*) can be expressed as

$$f = \frac{3\langle cos^2\phi \rangle - 1}{2} \quad (3.2)$$

$$\langle cos^2\phi \rangle = \frac{\sum_0^{\pi/2} I(\phi) cos^2\phi}{\sum_0^{\pi/2} I(\phi) sin\phi} \quad (3.3)$$

where $\phi$ and $I(\phi)$ represent azimuthal angle and intensity, respectively. The Herman's orientation factor (*f*) shows the orientation relative to direction of interest, where 0 indicates random orientation, -0.5 perpendicular orientation, and 1 parallel alignment to the surface. The characterisation of orientation can be performed by Raman spectroscopy as well Figures 7e & 7f. Qualitative measurement can be made by polarised Raman, comparing the intensity of the peaks at a set angle. Most researchers use an azimuthal scan of XRD to determine the degree of orientation of the nanofiller within samples.

**4.2. Dispersion of filler**

A homogeneous dispersion of filler is crucial in polymer nanocomposites. Agglomeration may occur in poorly dispersed nanofiller creating micron-sized aggregates. Air may trap inside the aggregates to produce void within nanocomposite lowering its properties [241]. In solution mixing, lyophilic colloidal suspension is essential to create concentrated suspension. Lyophilic colloid occur when there is strong interaction between particles and solvent or dispersing agent. For example, negatively charged titanate nanosheets are more stable in aqueous solution of tetrabutylammonium ($TBA^+$) cations [242]. Positively charged nanosheets, such as Mg-Al layered double hydroxides, are more stable in formamide, since formamide has a carbonyl group and is highly polar [243]. Having polar oxygen-containing functional groups, graphene oxide is stable in water, DMF, NMP, THF, ethylene glycol [244]. The dispersion limit of the graphene oxide in water may reach up to 9 g $L^{-1}$ by addition of surfactant [245], which is significantly higher than CNT at 1 g $L^{-1}$ even after surface modification [246]. However, some commercially available polymers (e.g., polypropylene, polyethylene, polystyrene) cannot be dissolved in common polar solvent. Treatments are needed to alter the surface of nanofiller stabilising its dispersion in polymer solution.

Chemical treatments are often used to improve the dispersion of filler [247]. The surface of graphene oxide can be modified (e.g., alkylated [248], coated with other polymer [249]) to be hydrophobic preventing agglomeration in non-polar solvent such as xylene and hexane. *In-situ* polymerisation of polypropylene can be conducted on the surface of graphene oxide by modifying the surface of graphene oxide with Ziegler-Natta catalyst [250]. Filler can also be dispersed homogeneously, even in polymer melt by surface modification. Alkylammonium modified montmorillonite clay alters its surface to organophilic thus polystyrene melt is able to intercalate into clay layers [251]. The interaction between filler and polymer must be controlled. Liu, *et al.* [252] has studied the correlation between dispersion and filler-polymer interaction using coarse-grained molecular dynamics. It was found that moderate interaction between filler and polymer creates optimum dispersion of filler. At low interaction, filler tend to attract each other producing agglomerates while strong interaction between polymer and filler creating flocculates due to strong polymer adsorption to its neighbouring filler. Physical treatment may also help dispersion of filler. However, one must consider that physical treatment such as ultrasonication, shear mixing, and ball milling might damage the filler thus optimum duration of treatment need to be studied. Besides the chemical and physical treatments of the filler, recent studies provide new approaches by using the interaction of filler with air bubble or oil [253] as well as synergetic effect by additional constituents such as clay [254], graphene oxide [255], and nanocellulose [134]. By tuning the pH of the solution, the graphene oxide (GO) can acted as surfactant with hydrophilic edges and a more hydrophobic basal plane enabling GO absorption on the interface of air-water and liquid-liquid [253]. The amphi-philicity of GO were created due to the degree of ionization of the edge -COOH groups which affected by pH. Strong affinity between clay and carbon nanomaterials (e.g., CNT, graphene) [256,257] prevented aggregation of each constituent and created network that lower the electrical percolation threshold [254]. Such synergetic interaction also occurred between graphene oxide and carbon nanotube [255] as well as the nanocellulose with carbon nanomaterials [134]. Clay, graphene oxide, and nanocellulose have high stability in aqueous medium helping carbon nanomaterials to disperse in aqueous solution. One must note that nanocelluloses were able to increase the dispersion limit of the carbon nanotube up to 60 wt% and 75 wt% in water by incorporation of cellulose nanofibrils (CNF) and cellulose nanocrystals (CNC), respectively [134]. It was found that high surface charge density of nanocellulose (1400 $\mu$equiv $g^{-1}$) was necessary to create sufficient interaction with carbon nanomaterials producing stable dispersion. Such strong interaction was also create synergetic effect on the mechanical properties of nacre inspired graphene/CNC with 765 ± 43 MPa and 15.64 ± 2.20 MJ $m^{-3}$ for tensile strength and toughness, respectively [258].

Although treatments may disperse the fillers within solution, the filler content within polymer is limited. Viscosity of the polymer solution is usually increased to form solid-like viscoelastic behaviour by incorporation of nanofiller (e.g., CNT, graphene) which is undesirable for solution mixing [259]. However, anomalous behaviour has recently been observed [153,154,260–263]. Slip between silica nanospheres and polymer occurred depend on particles size and polymer chain size, decreasing the bulk viscosity of polymer nanocomposite [261]. Strengthening interaction between silica nanospheres and polypropylene via grafting vinyl triethoxysilane yielded higher viscosity of nanocomposite than pure propylene [154]. In some cases, nanoparticles may selectively attract high molar mass fraction of the polymer leading to reduction in entanglement in polymer chain thus lowering apparent dynamic viscosity [153–155]. Better understanding of nanofiller-polymer matrix interaction may help processing of polymer nanocomposite with high filler content. Alternatively, the preform of nanofiller can be deployed to ensure the dispersion of nanofiller within polymer matrix yet the impregnation of polymer into the preform must be ensured, for example by controlling viscosity of the polymer and strong interaction between polymer and filler. The nacre bio-mimetic approach also produce a good dispersion of polymer since the polymer can be uniformly absorbed or attached to the dispersed filler in solution before solvent removal to create "brick" and "mortar" structure. The polyethylene imine (PEI) modified mica enabled the mica to absorbed chitosan exhibiting uniform dispersion and exfoliation of nanosheets up to 60 wt% of filler [264]. This composite exhibited 259 MPa and 16.2 GPa for tensile strength and modulus, respectively, while maintaining the transparency above 60% visible light transmission at 25 micron thick.

### 4.3. Interfacial bonding between filler and polymer matrix

Interface between nanofiller and polymer is important because it dictates the stress transfer efficiency in polymer nanocomposite. It can be strengthened by introducing functional groups to the nanofiller. It was established that there are two common method to functionalised nanostructures which is covalent (defect & side-wall functionalisation) and non-covalent (π–π interactions, hydrogen & ionic bonding) [265]. The covalent chemical bond between nanofiller and polymer exerts high interfacial strength improving the load and heat transfer on the interfaces of the composites depending on the moieties of the linkages. For example, butyl groups are more effective in reducing interfacial thermal resistance compared to carboxyl and hydroxyl according to molecular dynamic simulation [266]. However, the covalent modification of the carbon surface induces the hybridisation from $sp^2$ to $sp^3$ reducing conjugated bonds and lowering its electrical properties [267]. The non-covalent functionalisation can be used as alternative, which may improve the composite strength while maintaining the electrical properties of the nanofiller. The strength of non-covalent functionalisation can be determined by the combined effect of attractive forces (electrostatic, dispersive, and inductive interactions) and repulsive forces (exchange repulsion). In carbonaceous materials, the non-covalent functionalisation depends on the $\pi$ interactions (e.g., H-$\pi$, $\pi$-$\pi$, cation-$\pi$, and anion-$\pi$). Depending on the stiffness of the polymer backbone, the aromatic conjugated polymer chains may create a helical wrap around the lateral side of the carbon nanotubes through $\pi$-$\pi$ interactions. The wrapping mechanism occurs when the number of polymer units interacting with the nanotube surface at the expense of some torsional energy necessary to distort are sufficient. According to Jorge *et al.* [268], the binding energy is stronger for more flexible polymer chain. However, flexible polymer with bulky aromatic side groups tend to make intrachain coiling rather than wrapping to the nanotube [269]. This polymer wrapping also depends on the chirality of carbon nanotubes hence it can be used to separate semiconductor CNT from metallic CNT up to 99.85% purity [270].

Generally, the stronger interface between nanofiller and polymer increases the strength of the composite but lowering its toughness due to the brittle nature of the filler and the absence of crack deflection at the interface [271]. Toughness is the ability of a material to absorb energy and plastically deform which combination of the strength to bear a load and ductility. The composite with weak interfaces usually has high ductility due to the occurrence of crack deflection and crack bridging when deformation occurs. The combination of covalent and non-covalent interaction may create synergy of strong and weak interfaces improving toughness and its strength simultaneously. Synergetic effect of covalent and non-covalent functionalisation has been studied for improvement of polymer nanocomposite mechanical properties [21]. Figure 9 shows recent development of synergetic effect in nacre graphene/polymer composite. The highest toughness and tensile strength resulted from the synergistic effect of covalent bonding and non-covalent bonding (e.g., $\pi$-$\pi$ interactions and ionic bonding). To balance the strong covalent bonding, the non-covalent (e.g., $\pi$-$\pi$ interactions and ionic bonding) is crucial to dissipate the energy allowing plastic deformation. Graphene fibres usually have higher toughness than graphene films due to the reorientation of graphene during tensile testing absorbing deformation energy [272]. Bending and wrinkling of graphene may occur during the spinning process thus strong interaction between graphene and polymer is needed to engineer the defect of graphene [273]. The ternary composites also exhibit the synergetic effects by the interaction of two building blocks (e.g., graphene-CNT, graphene-$MoS_2$, graphene-MMT) which inspired by nature's hierarchical materials such as nacre with nanofibrillar chitin and aragonite calcium carbonate platelets. Natural hierarchical structures intrigues more researchers to synthesise man-made hierarchical structures such as tube-in-tube structures [274–276] and dots in tube [277], which also increased its functionality.

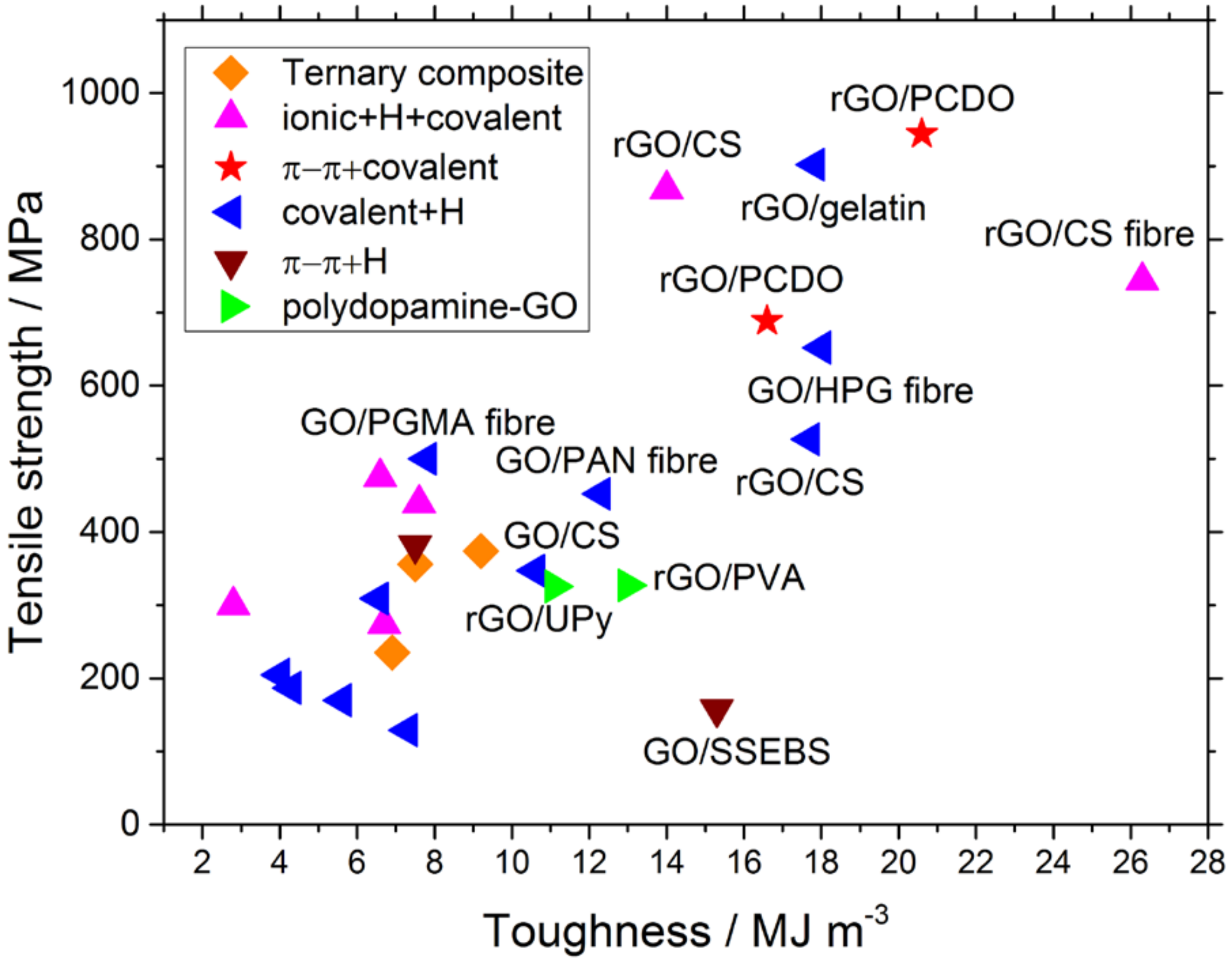

Figure 9. Recent developments in nacre graphene/polymer composite strength and toughness including graphene/polymer fibre [272,278–280], and its crack deflection strategy such as ternary composite [180,187,192], combination of ionic, hydrogen, and covalent bonding [181,184,192,193,198], $\pi$-$\pi$ interaction and covalent bonding [179], $\pi$-$\pi$ interaction and hydrogen bonding [185,190], polydopamine capped GO [188,189], and covalent and hydrogen bonding [182,183,186,194–197]

## 5. Conclusions and potential applications

While polymer nanocomposites of high filler content promise many potential applications, several improvements are needed to further maximise its potential. This chapter discuss the existing and future applications of these polymer nanocomposites which are shown in Figure 10.

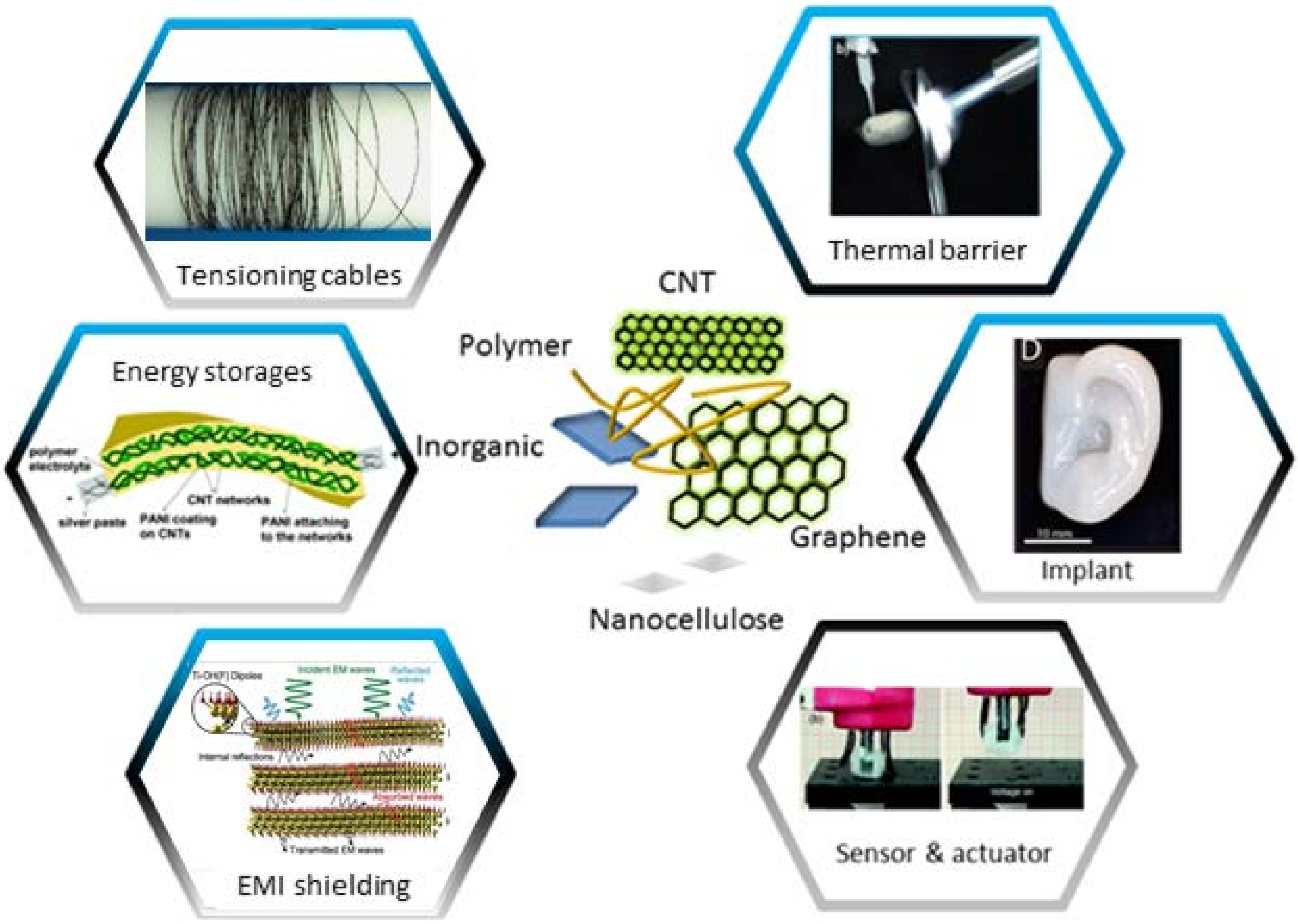

Figure 10**.** Existing and future application of high filler content of polymer nanocomposites. The images of energy storage [281], sensors & actuators [282], tissue engineering or implants [283], tensioning cables [284], thermal barriers [285], and electromagnetic interference (EMI) shielding [286] were taken from references. Adapted from Ref. [281, 283, 284, 285] with permission from © American Chemical Society. Reproduced from Ref. [282] with permission from The Royal Society of Chemistry. From Ref. [286]. Reprinted with permission from AAAS.

### 5.1. Existing applications

#### 5.1.1. Tensioning cables

The combination of nanostructural particles in polymer binder has been optimised for structural materials, heat barriers, adhesives, implants, and electronics devices. Recent advances in CNT and graphene fibre development, which can achieve the modulus of 330 GPa [70] and 282 GPa [54], respectively, are comparable with PAN based carbon fibre at 230-290 GPa [48] attracting researchers to develop its potential for structural composites. Due to their small diameter, CNT fibres have higher surface area than carbon fibre, which potentially offers improved load transfer between filler and polymer. On the other hand, the poor penetration of polymer into the CNT fibres yields voids in the composites lowering its mechanical properties. The approaches to impregnate CNT fibres with polymer can be categorised by the time of polymer impregnation, which is before, during, and after CNT spinning. The CNT/BMI composite fibre, which the polymer impregnation occurred before fibre spinning, yielded the highest mechanical properties with the tensile strength and modulus of 4.5–6.94 GPa and 232–315 GPa, respectively [84]. In additive manufacturing, the CNT/polyetherimide composite fibre was deployed as feedstock [287]. The absence of additional polymer made the CNT/polyetherimide fibre more versatile than other conductive feedstock (e.g., conductive silver, chopped carbon fibre, CNT, and graphene). A group of researchers studied the utilisation of CNT/epoxy composite fibre for tension dominated applications such as composite overwrapped pressure vessels (COPVs) [288], which usually used for maintaining pressure of cryogenic tank in space vehicle. Wet winding of CNT/epoxy composite fibre over aluminium rings resulted in a 209 % increase the hoop tensile properties and 10.8 % increase in weight relative to the bare ring, which was similar reinforcement as carbon fibre composites.

### 5.1.2. Electromagnetic interference (EMI) shielding

The electromagnetic interference (EMI) may disturb the signal of electronic devices and harmful for human health thus it must be shielded to prevent leakage of electromagnetic field radiation [289]. As the device become smaller in recent years, shielding with minimal thickness and lightweight is desirable. The EMI shielding material should be a conductive and/or magnetic material to reflect, absorb, or scatter the electromagnetic radiation [286]. Although metal has relatively high EMI shielding effectiveness (EMI SE), polymer nanocomposites are getting more attention due to its lightweight, high process-ability and non-susceptible to corrosion. The recent development of EMI shielding materials is provided in Figure 11.

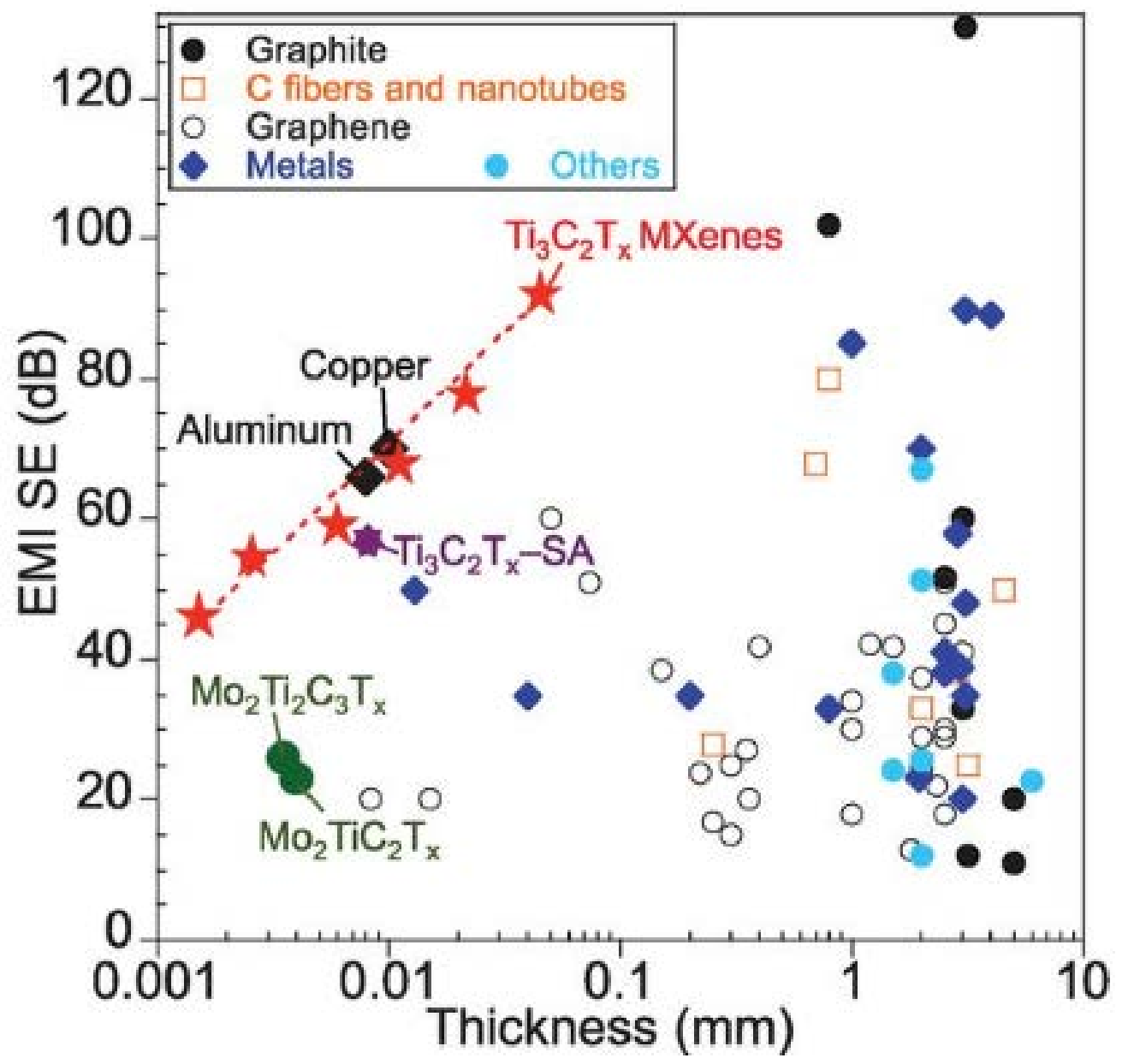

Figure 11. Comparison of electromagnetic interference shielding effectiveness (EMI SE) of various materials with its thickness [286]. From Ref. [286]. Reprinted with permission from AAAS.

The emerging materials, 2D transition metal carbides (MXenes), were exhibited excellent performance of EMI SE with minimum thickness. The EMI SE reached up to 92 dB for 45 micron pure $Ti_3C_2T_x$ MXenes film (T stands for surface-terminating functional groups like -F, -OH, -O) [286]. It can also be incorporated within polymer using nacre bio-mimetic synthesis method to induce flexibility. The EMI SE of $Ti_3C_2T_x$/sodium alginate (90:10) wt% composite was 57 dB at 8 micron, which was comparable to metal. However, the Young's modulus of $Ti_3C_2T_x$ MXenes was only 0.33 ± 0.03 TPa [290], which was below graphene at ≈1 TPa [17]. The rGO/calcium alginate (95:5) wt% exhibited 25.7 dB at 12 micron with 118 MPa and 4.6 $MJ.m^{-3}$ for tensile strength and toughness, respectively [291]. Alternatively, ternary composite of graphene, MXenes, and polymer can be synthesised for EMI shielding in the future.

### 5.1.3. Energy storage

According to Thomas *et al.* [292], an electrode for energy storage should have certain properties such as high surface area, mechanical and chemical stability, electrical and ionic conductivity to fabricate high performance electrode. The excellent properties of high filler content polymer nanocomposites enabled its utilisation for supercapacitors [293,294]. Due to its conductivity and chemical stability, conducting polymers (e.g., polypyrrole (PPy), polyaniline (PANI), poly(3,4-ethylenedioxythiophene) (PEDOT)) become a suitable choice to be used as polymer binder for energy storage [295]. Recent development of nanofiller/conducting polymer composite for energy storage is provided in Table 4.

Table 4. Recent development of nanofiller/conducting polymer composites for supercapacitors

| Electrode materials | Electrolyte | Method | Specific or volumetric capacitance | Capacitance retention |
|---|---|---|---|---|
| rGO/PANI (3:2) layer [296] | 1 M $H_2SO_4$ | Layer-by layer assembly | 736 F $cm^{-3}$ at 10 mV $s^{-1}$ | 86.6% after 5000 cycles |
| CNT/PANI (44:56) wt% [297] | 1 M $H_2SO_4$ | *In-situ* electro-polymerisation | 359 F $g^{-1}$ at 4.95 A $g^{-1}$ | 80% after 5000 cycles |
| f-CNT/PANI (65:35) wt% [298] | 0.5 M $H_2SO_4$ | *In-situ* chemical polymerisation | 1065 F $g^{-1}$ at 1 A $g^{-1}$ | 92.2% after 1000 cycles |
| rGO/CNT/PANI (1:1:1) mass ratio [299] | 0.1 M $NaNO_3$ | Filler mixing | 312.5 F $g^{-1}$ at 0.1 A $g^{-1}$ | 94% after 25 cycles |
| rGO/CNT/PANI (60:9:31) wt% [300] | 1 M HCl | Filler mixing | 569 F $g^{-1}$ at 0.1 A $g^{-1}$ | 96% after 5000 cycles |
| $rGO/MnO_2/PANI$ (10:70:20) wt% [301] | 1 M $Na_2SO_4$ | *In-situ* chemical polymerisation | 512 F $g^{-1}$ at 0.25 A $g^{-1}$ | 97% after 5000 cycles |
| $rGO/MnO_2/PANI$ (15:70:15) wt% [302] | 1 M $H_2SO_4$ | *In-situ* chemical polymerisation at oil-water interface | 800.1 F $g^{-1}$ at 0.4 A $g^{-1}$ | 71% after 800 cycles |
| f-CNT/$MnO_2$/PPy (35:21:44) wt% [303] | 0.5 M $Na_2SO_4$ | *In-situ* chemical polymerisation | 268 F $g^{-1}$ at 5 mV $s^{-1}$ | 90% after 5000 cycles |
| CNT/PPy fibre (49:51) wt% [304] | 1 M $H_2SO_4$ | *In-situ* electro-polymerisation | 350 F $g^{-1}$ at 1 A $g^{-1}$ | 88% after 5000 cycles |
| CNT/PPy fibre (50:50) wt% [305] | 1 M $H_2SO_4$ | *In-situ* electro-polymerisation | 302 F $g^{-1}$ at 1 A $g^{-1}$ | 96% after 5000 cycles |
| $Ti_3C_2T_x$/PPy (2:1) mass ratio [306] | 1 M $H_2SO_4$ | *In-situ* chemical polymerisation | 1000 F $cm^{-3}$ or 416 F $g^{-1}$ at 5 mV $s^{-1}$ | 92% after 25000 cycles |
| $MoS_2$/PPy (25.8:1) mass ratio [307] | 1 M KCl | *In-situ* chemical polymerisation | 695 F $g^{-1}$ at 0.5 A $g^{-1}$ | 85% after 4000 cycles |
| $V_2O_5$/PPy (50:50) wt% [308] | 0.5 M $K_2SO_4$ | *In-situ* chemical polymerisation | 308 F $g^{-1}$ at 0.1 A $g^{-1}$ | 95% after 10000 cycles |

f-CNT = functionalised CNT

For scaffold impregnation method, *in-situ* polymerisation was the common synthesis method to fabricate conducting polymer based composite. The polymerisation can be conducted either by the chemical reactions of monomer, dopant and oxidant at usually 0 $^oC$ or by electrochemical deposition of monomer and dopant [309]. The functionalisation can be performed on the nanofiller improving the adsorption and dispersion of monomers on the surface of the filler [298,303]. For filler mixing method, the hybrid nanofillers may increase the dispersion limit of the nanofiller. For example, carbon nanotube can be stabilised in polymer solution by the addition of graphene oxide enabling the fabrication of high filler content polymer composite [299,300]. The high filler content/conducting polymer composite exhibited promising result to be used as supercapacitor especially $Ti_3C_2T_x$ MXene/PPy, which showed the capacitance retention of 92% even after 25000 cycles [306]. However, the researches on MXenes supercapacitor are still limited.

**5.1.4. Sensors and actuators**

The performance of polymer nanocomposite strain sensor should be stable during stretching as well as durable to stretching, bending, and folding. As discussed in sections 3 and 4, the high filler content polymer composite provides high tensile strength, toughness, and electrical conductivity, which is desirable properties for strain sensor. For example, the nacre inspired $Ti_3C_2T_x$ MXenes/Ag nanowire/polydopamine composite with $Ni^{2+}$ ion crosslinker produced a stable relative resistance changes during 5000 stretch cycles of 0 to 60% strain with a sensitivity (gauge factor) of 1160.8 under 60% strain [310]. The non-covalent bonding (e.g., hydrogen bonding) of nacre composite can be altered by several stimuli such as photothermal [311], water [312], and alcohol [313,314] environments, which enable the composite to act as actuator to create certain shape. Driven by a humidity gradient, the moisture-responsive GO/chitosan demonstrated bending ability up to 180 $^{o}$, which was also able to lift objects 50 times heavier and transporting cargos 10 times heavier than itself [312]. By combination of folding and cutting, this shape memory properties may create a perpetual machine for biomedical tools and soft robotics [315].

**5.1.5. Thermal barriers**

For thermal barrier applications, the nacre inspired ceramic based nanosheets, which have high thermal resistance, is very promising. The excellent fire resistance of “brick-and-mortar” architecture of ceramic based polymer composites have been discussed in section 3.2. However, most of the nacre composites used flammable polymer such as polyvinyl alcohol or polyacrylic acid. The flame resistant of nacre can be further improved by choosing fire resistant polymer [285]. For example, the micron thick montmorillonite/polyanion composite had better fire resistant than montmorillonite/polyvinyl alcohol, which was able to protect silk cocoon, which positioned 8 mm behind the composite, from *ca.* 2000 $^{o}$C gas burner exposed to the other side of composite [285]. The high content of montmorillonite based composite can be deployed as fire resistant coating of textiles [316]. Das *et al.* [316] combined montmorillonite (MTM) with sodium carboxymethyl cellulose (CMC) with the composition of 40/60 wt% to create 1-5 wt% slurry for dip coating the pure cotton textile. The nacre biomimetic coating of MTM/CMC was able to protect the cotton from the torch flame (ca. 1750 °C) at nearly 90° angle and 130 mm away from the sample without observable shrinkage. These flame retardant properties were attributed to tortuous pathway of oxygen diffusion within “brick and mortar” architecture, which the oxygen barrier properties observed was as low as 0.022 $cm^3.mm.m^{-2}.day^{-1}.atm^{-1}$ at 50 % relative humidity [316]. The tortuous pathway of gas in nacre inspired polymer composites indicates its potential in gas barrier application.

### 5.1.6. Tissue engineering scaffolds or implants

High porosity, 3D foams can prove suitable as tissue engineering scaffolds. Graphene and carbon nanotubes are known to be biocompatible, aid neural growth, and support the adhesion and proliferation of bone cells [317,318]. Abarrategi *et al.* [161] synthesised the 3D carbon nanotube/chitosan foam scaffold by unidirectional freeze drying of dispersed CNT in chitosan solution. This technique was able to produce 3D foam with CNT content up to 89 wt%, which used to perform *in vitro* and *in vivo* evaluation for cell adhesion, viability and proliferation of C2C12 cell line (myoblastic mouse cell). The micron size porosity enabled the infiltration of cell into the pore channels and the cell adsorption onto the surface. Besides carbon based polymer composite, the nanocellulose based polymer composites have tremendous potential in medical applications [319]. Due to its biodegradable nature, biocompatibility, and low cytotoxicity, the nanocellulose based composites was used as bio-ink to fabricate 3D bio-printed composite hydrogel implant for cartilage tissue engineering applications [283,320]. The nanofibrillated cellulose (NFC), which made through enzymatic hydrolysis, mechanical shearing, and high-pressure homogenization, was combined with alginate to 3D bio-print human chondrocytes with high fidelity and stability with the optimum compressive stiffness achieved at 70:30 wt% ratio of NFC and alginate [283]. However, the nanofibrillated cellulose (NFC) obtained by TEMPO-oxidation was not stable as 3D bio-printed hydrogel due to water evaporation, requiring the addition of glycerine as high as 50 wt/vol% to stabilise the form at room temperature [321].

### 5.2. Future applications

The development of carbon based polymer nanocomposite has dominated the utilisation of high filler content composites in diverse applications from tensioning cables, EMI shielding to electrical devices due to its high mechanical [2,17] and electrical properties [15]. According to recent toxicity studies [322,323], the nanosize carbonaceous materials may be harmful for human respiratory system which causes pulmonary inflammation and toxicity. The nanocellulose, which has biodegradable nature and less toxic, can be used as alternative for implant or tissue engineering scaffold although it may become toxic at very high dose of nanocellulose [324]. Meanwhile, the high thermal resistance of inorganic filler aids the composite acting as a thermal barrier. This review provides several technique to synthesis high filler content composite, which may open new possibilities for the application of inorganic and nanocellulose based polymer composites. The 3D ternary composite foam can be synthesised to absorb and degrade dye or organic pollutant [325], in which the hybrid strategy to disperse the carbon nanomaterials can be utilised. The non-enzymatic non-precious metal saccharide (e.g., glucose, fructose) sensor may benefit from the combination of metal oxides and conducting polymer. Some transition metal and its metal oxides, especially Cu-, Ni-, Co- based oxides, have attracted attention as glucose oxidation catalysts of high sensitivity [326]. Several materials such as $TiO_2$ nanosheets [327,328] and conducting polymer (e.g., polyaniline) [329] can be used to bind the boronic acid moieties, which have a strong affinity towards polyols such as carbohydrates derivatives, making it a promising for saccharide and organic analysis. By combining metal oxide with conducting polymer and boronic acid, the charge carriers of the metal oxide will be improved and boronic acid functionalities may induce selective detection of saccharides. Moreover, the metal oxide gas sensor could be benefitted by the incorporation of conducting polymer as well to obtain reasonable conductance and speed of response and recovery time at lower operational temperature. One must note the dopant greatly influence the sensitivity of conducting polymer gas sensor [330]. Careful selection of dopant and the conducting polymer should be performed to fabricate high quality metal oxide/conducting polymer composite gas sensor.

Despite the utilisation of nanocellulose-based hydrogel composite, the use of nanocellulose as filler in rigid polymer is limited due to poor dispersion of nanocellulose within the rigid polymer at high nanocellulose content. The mechanical properties of the nanocellulose/polymer composites was lower than its pure polymer at more than 20 vol% of nanocellullose [26]. Several attempts have been made to fabricate 1D nanocellulose yarn by wet and dry spinning of nanofibrillated cellulose (NFC) and regenerated cellulose [331]. The resulting 1D yarn of nanocellulose yielded a Young’s modulus with the range of 8-23 GPa, while the cellulose crystal itself was estimated to be up to 160 GPa in the longitudinal direction [332] and 8−57 GPa in the transverse direction [333]. Further studies are needed to optimise the NFC aspect ratio, solid content and shear rate of the spinneret, as well as additional drawing and stretching processes. The impregnation of polymer into the nanofibrillated cellulose (NFC) yarn may also facilitate stress transfer within fibres and imbues the NFC yarn with additional functionalities for example as bio-based structural material.

Generally, the nacre inspired nanocellulose composites involves additional carbon or clay based nanosheets as a "brick" structure to form "brick and mortar" architecture. There have been several attempts to synthesise cellulose nanosheets by a top-down approach (e.g., ball milling) [334] and bottom-up approach (e.g., polymer capped nanosheet synthesis) [335]. The cellulose powders from cotton linters have been subjected to ball-mill with 10 mm zirconia balls in dry condition or with liquid such as water or PDMS silicone oil. The affinity of water and PDMS to specific plane in nanocellulose affected the resulting morphology, which are nanofibres for water and nanosheets for PDMS. The nanocelluloses were deformed with capping at (200) lattice plane forming nanosheets with 1-5 micron in lateral dimension and thickness around 5 nm. The polymer capping strategy can be deployed as well in enzymatic synthesis of cellulose via in vitro cellodextrin phosphorylase (CDP) mediated polymerization of α-D-glucose 1-phosphate (αG1P) using glucose or cellobiose as a primer. The vinyl-cellulose nanosheets created by the enzymatic synthesis exhibits elongated nanosheets with 5 nm thickness, ≈200 nm width and ≈500 nm length. These cellulose nanosheets could be used as "brick" in nacre inspired polymer composite. Although emerging materials, such as MXenes, have shown potential use in many applications, studies of high content MXenes/polymer composites are still limited. Approaches to synthesis high content polymer nanocomposite could realise novel applications of these emerging materials as alternative to CNT and graphene. A maximum potential reinforcement of nanoscale filler is expected without limitation in orientation, dispersion, and interfacial problems.

**Acknowledgements**

This research was partially supported by World Class University Program ITB decree number 237/SK/I1.B02/2018 and Indonesia Endowment Fund for Education (LPDP).

## Conflicts of interest

There are no conflicts to declare